\documentclass[lettersize,journal]{IEEEtran}
\usepackage{amsmath,amsfonts}

\usepackage{array}
\usepackage[ruled,vlined]{algorithm2e}
\usepackage[caption=false,font=normalsize,labelfont=sf,textfont=sf]{subfig}

\renewcommand{\vec}[1]{\boldsymbol{#1}}
\allowdisplaybreaks[4]

\usepackage[colorlinks=true,urlcolor=blue]{hyperref}
\usepackage{arydshln} 
\usepackage{textcomp}
\usepackage{stfloats}
\usepackage{url}
\usepackage{verbatim}
\usepackage{graphicx}
\usepackage{cite}
\usepackage[utf8]{inputenc}
\usepackage{indentfirst}
\usepackage{graphicx}
\usepackage{amssymb}
\usepackage{tabularx}
\usepackage{bm}
\usepackage{color}

\allowdisplaybreaks[4]

\title{A Fast Task Offloading Optimization Framework for IRS-Assisted Multi-Access Edge Computing System}
\begin{document}
\author{Jianqiu Wu,~Zhongyi Yu, Jianxiong Guo,~\IEEEmembership{Member,~IEEE}, Zhiqing Tang, Tian Wang, and Weijia Jia,~\IEEEmembership{Fellow,~IEEE}
    \thanks{Jianqiu Wu and Zhongyi Yu are with the Guangdong Key Lab of AI and Multi-Modal Data Processing, Department of Computer Science, BNU-HKBU United International College, Zhuhai 519087, China. (E-mail: jqwuhelen@qq.com; zhongyicst@gmail.com)

    Jianxiong Guo, Tian Wang, and Weijia Jia are with the Advanced Institute of Natural Sciences, Beijing Normal University, Zhuhai 519087, China, and also with the Guangdong Key Lab of AI and Multi-Modal Data Processing, BNU-HKBU United International College, Zhuhai 519087, China. (E-mail: jianxiongguo@bnu.edu.cn; cs\_tianwang@163.com; jiawj@bnu.edu.cn)

    Zhiqing Tang is with the Advanced Institute of Natural Sciences, Beijing Normal University, Zhuhai 519087, China. (E-mail: zhiqingtang@bnu.edu.cn)
    
    \textit{(Corresponding author: Jianxiong Guo.)}
	}
    \thanks{Manuscript received April xxxx; revised August xxxx.}}

\markboth{Journal of \LaTeX\ Class Files,~Vol.~xx, No.~xx, July~2023}%
{Shell \MakeLowercase{\textit{et al.}}: Bare Demo of IEEEtran.cls for IEEE Journals}

\maketitle

\begin{abstract}
Terahertz communication networks and intelligent reflecting surfaces exhibit significant potential in advancing wireless networks, particularly within the domain of aerial-based multi-access edge computing systems. These technologies enable efficient offloading of computational tasks from user electronic devices to Unmanned Aerial Vehicles or local execution. For the generation of high-quality task-offloading allocations, conventional numerical optimization methods often struggle to solve challenging combinatorial optimization problems within the limited channel coherence time, thereby failing to respond quickly to dynamic changes in system conditions. To address this challenge, we propose a deep learning-based optimization framework called Iterative Order-Preserving policy Optimization (IOPO), which enables the generation of energy-efficient task-offloading decisions within milliseconds. Unlike exhaustive search methods, IOPO provides continuous updates to the offloading decisions without resorting to exhaustive search, resulting in accelerated convergence and reduced computational complexity, particularly when dealing with complex problems characterized by extensive solution spaces. Experimental results demonstrate that the proposed framework can generate energy-efficient task-offloading decisions within a very short time period, outperforming other benchmark methods.

\end{abstract}
\section{Introduction}

The widespread adoption of smart personal devices and web services has led to a significant surge in the demand for computational resources and high-speed transmission networks. In order to alleviate the computational burden on user devices and enhance user experience, Multi-access Edge Computing (MEC) systems have emerged as a promising solution. These systems enable the offloading of computationally intensive tasks to edge servers with robust computational power. However, there are situations where the existing ground-based MEC infrastructures may be inadequate to meet the escalating computational demand. To overcome this limitation, the integration of Unmanned Aerial Vehicles (UAVs) into the MEC system has been proposed \cite{hassan2021UAV, hua2021uav, tun2020energy}. Equipped with MEC servers and endowed with flexible mobility, UAVs can be dynamically deployed to locations where the computational power falls short of fulfilling the demand. Nonetheless, the limited battery capacity of UAVs presents a challenge for the efficient operation of UAV-enabled MEC systems. Therefore, the meticulous design of offloading decisions for UAV-enabled MEC systems is crucial in optimizing system performance and ensuring overall system reliability.

To meet the increasing demand for high-speed data transmission, terahertz (THz) networks and intelligent reflecting surfaces (IRSs) have emerged as highly promising technologies. THz networks leverage their remarkable attributes of vast available bandwidths and extremely short wavelengths, thus holding immense potential to support terabit-per-second data transmission and ultra-fast communication. Moreover, previous studies \cite{IRSPrinciples, Sum-RateTHz, chen2019sum-rate, 2020THzRisk, chaccour2020risk, Yijin2021, li2020reconfigurable} demonstrate the crucial role played by IRS in augmenting wireless communication performance and network transmission speed. By adroitly manipulating the reflecting elements within IRS, IRS offers the capability to optimize signal strength, extend coverage, and enhance the overall system capacity.

\begin{figure}[!t]
    \includegraphics[width=\linewidth]{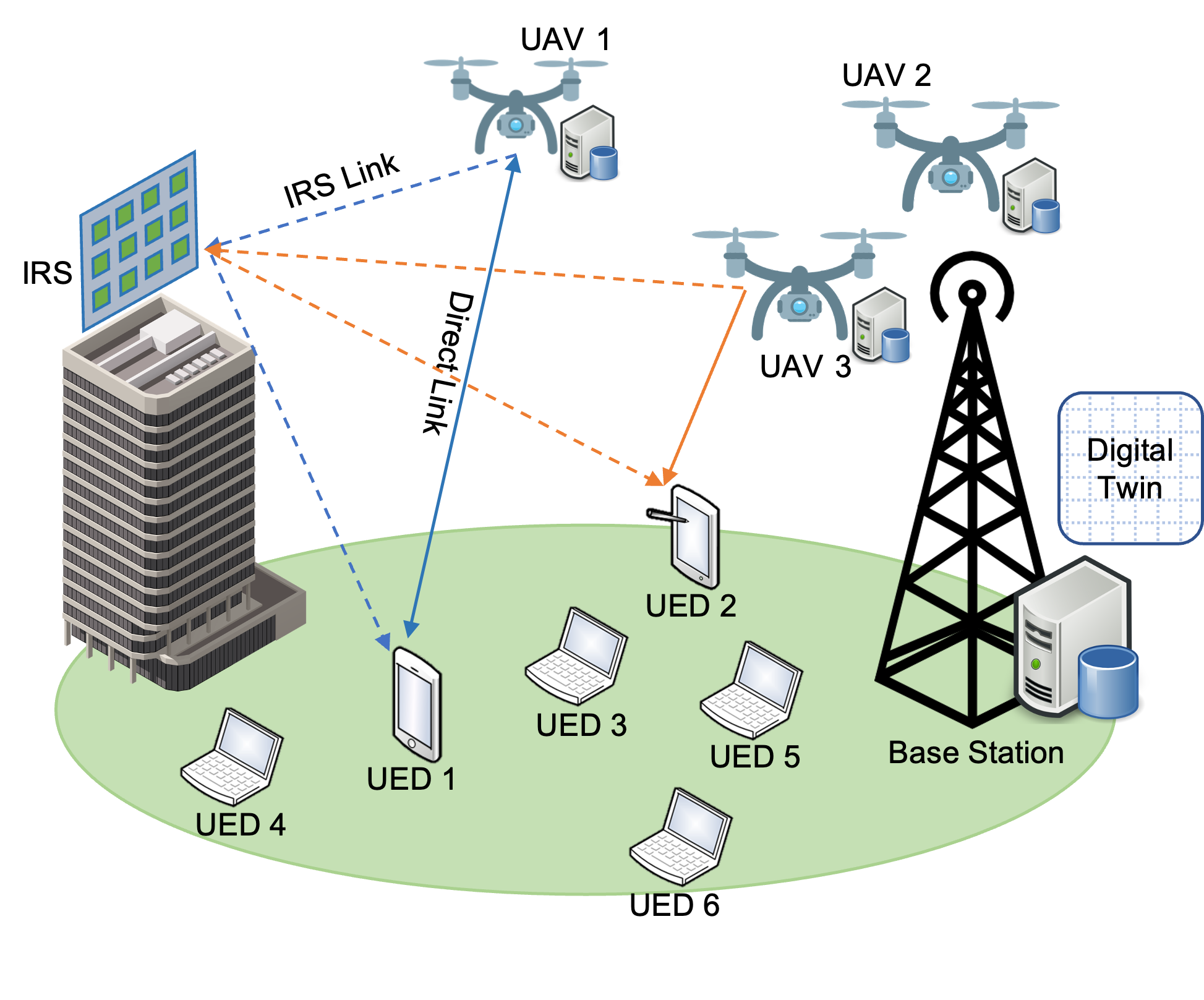}
    \caption{{The proposed MEC system.}}
    \label{fig:phy_sys}
\end{figure}

In this study, we aim to enhance MEC systems by leveraging the advantages of the IRS and the THz communication network. Specifically, we focus on optimizing task offloading allocations within an IRS-assisted MEC system operating in the THz network to minimize overall energy consumption. The proposed MEC system, illustrated in Figure 1, comprises multiple UEDs, a fleet of UAVs, and an IRS responsible for enhancing information transmission speed. In previous studies, \cite{Sum-RateTHz, chen2019sum-rate, 2020THzRisk, chaccour2020risk} introduce to enhance THz network communication with the incorporation of IRS but neglects the modeling of MEC systems within the context of THz networks. Additionally, while \cite{Yijin2021, li2020reconfigurable} propose IRS phase shift optimization techniques to improve system efficiency and reduce energy consumption, they do not specifically address the task offloading optimization problems in their respective systems. Regarding the optimization of offloading strategies, a large number of models have been introduced in \cite{2023Jing,2023Wang,2020DROO, 2019Dong, 2018Chen, 2019min, jiang2020,into_1,into_2,into_3,into_4,into_5,into_6,into_7,into_8}. However, these MEC systems overlook the simultaneous incorporation of IRS and UAVs and operate solely within the 5G network paradigm, disregarding the distinctive characteristics of THz communication networks, such as the THz communication model and the channel fading characteristics during signal transmission. As a closely related work to this study, \cite{2022Park} considers the inclusion of IRS and UAVs, and investigates the allocation of network sub-band and computational resources, as well as the optimization of IRS phase shifts in the context of THz networks. However, the studied system is not designed to address the MEC task offloading problem and only involves a single UAV, overlooking the complexities that arise in systems with multiple UAVs and a substantial number of users. As a result, task offloading allocations in an IRS-assisted multi-UAV MEC system operating within the THz network remains largely unexplored. 

To overcome these limitations, we begin by integrating IRS and UAVs into the MEC system and deploying this system within the THz communication network. Subsequently, we formulate the operation of this IRS-assisted multi-UAV MEC system within the THz network, taking into account the distinctive characteristics and challenges inherent to the THz network environment. Furthermore, we propose a novel deep learning framework named \textbf{Iterative Order-preserving Policy Optimization (IOPO)}. This framework effectively determines energy-efficient task offloading allocations for the MEC system and optimizes the phase shift configurations of the IRS. We extensively evaluate the performance of the proposed IOPO framework through numerical studies. Experimental results demonstrate that the IOPO framework surpasses baseline approaches in minimizing system energy costs and ensuring the timely completion of user tasks. Furthermore, experimental results indicate that IOPO is capable of generating optimal offloading allocations while adhering to the defined constraints. Our source code can be found in \url{https://github.com/UIC-JQ/IOPO}. The contributions of this paper can be summarized as follows.
\begin{itemize}
    \item We present a novel MEC system tailored for operation on the THz communication network. The proposed MEC system is equipped with an IRS, which plays a crucial role in enhancing communication performance within the network. Additionally, the system is designed to accommodate multiple UAVs as well as multiple users. 
    
    \item In order to streamline the optimization process and improve the efficiency of the MEC system, we propose a deep learning framework named IOPO. IOPO is designed to jointly optimize offloading decisions of the multi-user multi-uav system and the phase shift of the IRS. As a result, IOPO eliminates the need for solving complex Mixed Integer Non-Linear Programming (MINLP) problems, which can be computationally demanding and time-consuming. 
    
    \item To facilitate the generation of high-quality offloading decisions, we equip IOPO with a novel policy exploration unit called \textbf{Order-Preserving Policy Optimization (OPPO)}, which is specifically designed to search for improved offloading decisions. Experimental results demonstrate the effectiveness of OPPO in discovering improved offloading decisions, even in scenarios with a vast solution space. Furthermore, results show that the integration of OPPO facilitates the convergence of IOPO towards optimal offloading decisions.
\end{itemize}

The rest of the paper is organized as follows. Section \ref{sec:related_work} provides a comprehensive review of previous studies on THz communication and the MEC system. In Section \ref{sec:system_model}, we introduce the proposed MEC system model and formulate the data communication within the THz network. Section \ref{sec:optimization_model} formulates the optimization problem aimed at minimizing the energy consumed in the MEC system. The design of the proposed IOPO framework is described in Section \ref{sec:iopo}. Experimental settings are presented in Section \ref{sec:exp_settings}, followed by a thorough analysis of the results in Section \ref{sec:exp_results}. Finally, Section \ref{sec:conclusion} concludes the paper by summarizing the key findings.

\section{Related Work}
\label{sec:related_work}
The integration of IRS in THz communication has been extensively studied in recent works \cite{Sum-RateTHz, chen2019sum-rate, 2020THzRisk, chaccour2020risk, Yijin2021, li2020reconfigurable}. In \cite{Sum-RateTHz, chen2019sum-rate}, the IRS is employed to maximize the sum-rate performance of THz communications. The studies conducted in \cite{2020THzRisk, chaccour2020risk} focus on utilizing the IRS to maintain reliable THz transmission. \cite{Yijin2021} introduces a comprehensive optimization framework that jointly optimizes the UAV trajectory, IRS phase adjustments, THz sub-band allocation, and power control. Moreover, \cite{li2020reconfigurable} proposes a joint optimization approach for the UAV's trajectory and the IRS's beamforming, aiming to enhance the overall system performance.

To generate offloading allocations for MEC systems, several studies employ machine learning algorithms. \cite{2020DROO,2019Dong} applies deep reinforcement learning techniques to determine optimal task offloading strategies in scenarios involving single or multiple access points (APs). \cite{2018Chen} considers factors such as channel state information, queue state information, and energy queue state, and introduces a deep Q-learning network to generate offloading decisions that minimize task execution costs. Similarly, in \cite{2019min}, a deep Q-learning network is proposed to maximize the computational performance of energy-harvesting MEC networks. \cite{jiang2020} proposes a deep learning based optimization approach to minimize the system energy consumption while optimizing the positions of ground vehicles and unmanned aerial vehicles along with the resource allocation in a hybrid mobile edge computing platform. Furthermore, \cite{2022Park} focuses on optimizing the phase shift of IRS, UAV computing resources, and sub-band allocation in a single UAV scenario. These works demonstrate the effectiveness of machine learning models in producing high-quality offloading strategies for MEC systems. 

While progress has been made in existing literature, the task offloading in an IRS-assisted multi-UAV MEC system operating within the THz network remains unexplored. Specifically, \cite{Sum-RateTHz, chen2019sum-rate, 2020THzRisk, chaccour2020risk} primarily focuses on enhancing THz network communication with IRS. However, they do not adequately address the modeling of MEC systems within the context of THz networks. Moreover, \cite{Yijin2021, li2020reconfigurable} introduce the utilization of IRS to improve the efficiency of MEC systems, but their systems do not tackle the optimization problems associated with task offloading. Furthermore, \cite{2020DROO, 2019Dong, 2018Chen, 2019min, jiang2020} leverages deep learning models to produce offloading decisions. However, their proposed systems operate within the 5G network paradigm, neglecting the distinctive characteristics of THz communication networks. Lastly, \cite{2022Park} investigates the allocation of network recourses and computational resources in the context of THz networks, taking into account the integration of IRS and UAVs. However, the studied system does not address the MEC task offloading problem and only involves a single UAV, thereby failing to model the complexities that arise in systems with multiple UAVs.

\section{System model}
\label{sec:system_model}
In this section, we first provide a detailed description of the components comprising the proposed MEC system and demonstrate how the MEC system operates in general. Following this, Section \ref{sec:sys_model_data_transmit} formulates the communication and data transmission between UAVs and users within the MEC system. Lastly, Section \ref{sec:sys_model_eng} introduces the steps for computing the total energy consumed in the MEC system.

\subsection{The Proposed MEC System}
Figure \ref{fig:phy_sys} presents the proposed multi-UAV multi-user MEC system designed for 6G THz communication networks. The system comprises a single IRS, $U$ users denoted as $\mathcal{U} = \{1, 2, \cdots, U\}$, and $M$ UAVs denoted as $\mathcal{M}= \{1, 2, \cdots, M\}$. Each user is equipped with a User Electronic Device (UED), which serves as a local computing server. Each UAV provides full-duplex communication services to users within a specific area and is equipped with a MEC server responsible for processing the tasks uploaded by users and transmitting the results through downlink transmission. For simplicity, we refer to the MEC server mounted on the UAV as the UAV itself. To alleviate the computational burden on the UEDs, the MEC servers are designed with higher computational capacity. This empowers users to make decisions regarding task offloading, choosing between offloading their computational tasks to one of the $M$ UAVs or executing them locally on their UEDs. Consequently, the task allocation for the entire MEC system can be represented by a $U\times (M+1)$ matrix, where $M+1$ signifies that users choose from $M$ UAVs and their local UEDs. Regarding the IRS, it is comprised of $K$ reflecting elements. By manipulating the phase shifts of these reflecting elements, the IRS can reconfigure wireless propagation channels in a highly efficient manner. This reconfiguration leads to significant improvements in both the overall propagation environment and the data transmission speed of the system. 

The proposed MEC system operates as follows: at a time frame $n$
within the system time $\mathcal{N} = \{1, 2, \cdots,n,\cdots, N\}$, users in the system held computational tasks that need to be processed. The primary objective is to utilize the available computational resources, such as UAVs and UEDs, to complete all users' tasks within an acceptable time while minimizing the total energy consumed during task processing. To achieve this objective, an offloading decision that allocates user tasks to the appropriate computational resources is required. Initially, the central server, located at the base station, collects the necessary information. Subsequently, the collected information is input into an offloading decision prediction model, which is discussed in detail in Section \ref{sec:iopo}. This model predicts an offloading allocation matrix denoted as $\vec{\beta}(n) \in \{0,1\}^{U\times(M+1)}$, where $U$ represents the number of users and $M$ represents the number of UAVs. For a given user $u$, $\beta_{u,m}(n) = 1$ indicates that the corresponding task is offloaded to UAV $m$ ($m\leq M$), and $\beta_{u, M+1}(n) = 1$ signifies that the task is processed locally on the user's UED. In the proposed system, we assume that when a task is offloaded to UAVs, it can only be offloaded to a single UAV at a time, prohibiting simultaneous offloading to multiple UAVs. This constraint is mathematically expressed as $\sum_{m=1}^{M}\beta_{u,m}(n) = 1$ for each user $u \in \mathcal{U}$. Following the generation of the  offloading decision, it undergoes simulation in a digital twin \cite{digital}, which serves as a real-world simulator. Subsequently, the phase shift of the IRS is optimized based on the generated offloading decision. Finally, the energy cost associated with the offloading decision is obtained from the digital twin and utilized to update the offloading decision prediction model as required.

\subsection{Data Transmission in the THz Network}
\label{sec:sys_model_data_transmit}
\begin{figure}[!t]
    \includegraphics[width=\linewidth]{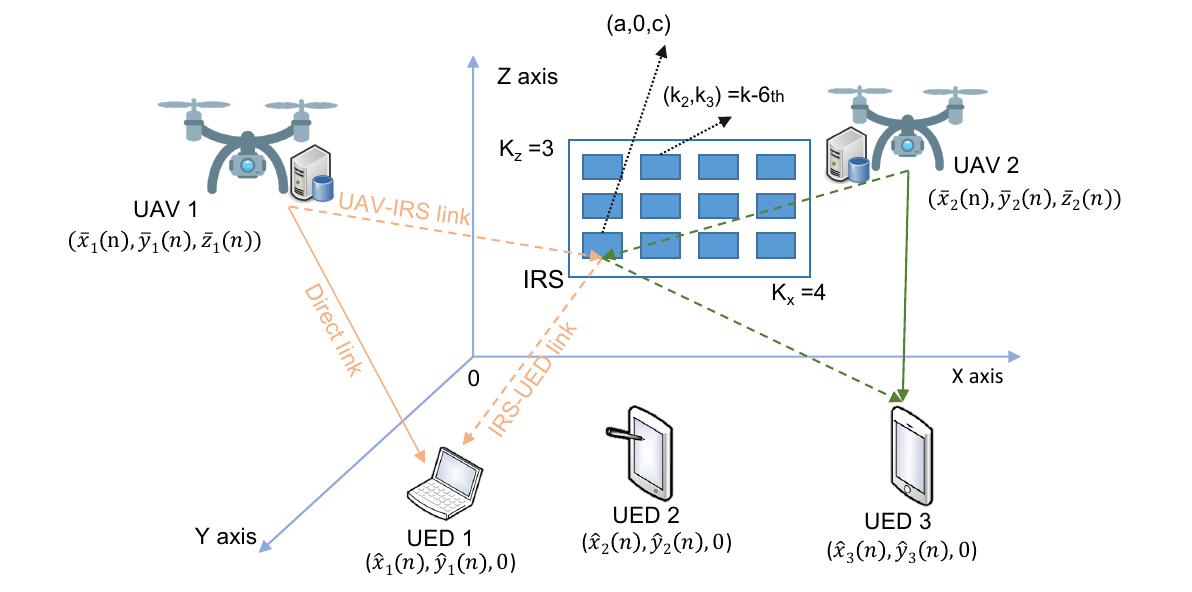}
    \caption{Data transmission in the proposed MEC system. User data can be directly transmitted from UEDs to UAVs or be redirected to UAVs from IRS.}
    \label{fig:transmission}
\end{figure}
In this section, we elucidate the data transmission within the THz network. As depicted in Figure \ref{fig:transmission}, at time frame $n$, there are two approaches for transmitting user data and tasks to UAVs: (i) direct transmission of user data from UEDs to UAVs, and (ii) redirection of user data to UAVs through the IRS. Both approaches are employed simultaneously in the system to facilitate efficient data transmission by the users. According to the Shannon theorem, the achievable throughput $R_{u,m}(n)$ for user $u$ to transmit data to the $m$-th UAV is determined as follows:
\begin{equation}
R_{u,m}(n) = B\log_2\left(1+\frac{p\left|h_{u,m}(n)+\hat{g}_{u,m}(n)\right|^2}{\sigma^2}\right),
\end{equation}
where $h_{u,m}(n)$ denotes the channel gain for direct data transmission and $\hat{g}_{u,m}(n)$ is the channel gain of transmitting data through the IRS. We assume that when multiple UEDs upload their tasks to UAVs simultaneously, the available wireless bandwidth is equally shared among them. Accordingly, B represents the channel bandwidth allocated to each UED. $p$ represents the transmission power provided by the base station and $\sigma^2$ is a Gaussian noise for modeling random noise that affects the communication. 

In the case of direct data transmission, given the coordinate of user $u$, denoted as $\hat{l}_u(n)=(\hat x_u(n),\hat y_u(n),0)^T$ and the coordinate of the $m$-th UAV, denoted as $\bar{l}_{m}(n) =(\bar x_{m}(n), \bar y_{m}(n), \bar z_m(n))^T$, the euclidean distance $d_{u,m}(n)$ between them can be formulated as:
\begin{equation}
    d_{u,m}(n) = \sqrt{(\bar x_m(n)-\hat x_u(n))^2+(\bar y_m(n)-\hat y_u(n))^2 + \bar z_m^2(n)}.\nonumber
\end{equation}

Given the distance $d_{u,m}(n)$, the channel gain for direct transmission $h_{u,m}(n)$ is defined as follows:
\begin{align}
h_{u,m}(n)&=\left(\frac{c}{4\pi f d_{u,m}(n)}\right) \cdot \nonumber\\
&\exp\left(\frac{-j2\pi f d_{u,m}(n)}{c}+\frac{-K(f)\,d_{u,m}(n)}{2}\right),
\end{align}
where $c$ represents the speed of light, $f$ denotes the frequency of the sub-band, $j$ is the imaginary unit, and $K(f)$ represents the absorption coefficient of the transmission medium.

In the context of data transmission through an IRS, the IRS acts as an intermediary that receives data from the data-sending device and subsequently reflects the data to the receiver. As depicted in Figure \ref{fig:transmission}, the IRS is situated on the X-Z plane and comprises a total of $K = K_x \cdot K_z$ reflecting elements. $K_x$ and $K_z$ represent the quantities of reflecting elements along the X-axis and Z-axis, respectively. The coordinates of the reflecting elements in the IRS are determined based on the position of the first reflecting element, denoted as $\tilde{l}_1 = (a, 0, c)^T$, which is located at the lower-left corner of the IRS. Accordingly, the coordinates of the $k$-th reflecting element ($k = k_z+(k_x-1)K_z$), denoted as $\tilde{l}_k$, can be calculated using the following expression:
\begin{equation}
\tilde{l}_k = \bigg(a + (k_x-1)\delta_x, \, 0, \, c + (k_z-1)\delta_z \bigg)^T, \nonumber
\end{equation}
where $k_x$ and $k_z$ represent the indices of the reflecting element along the X-axis and Z-axis, respectively. $\delta_x$ and $\delta_z$ denote the gaps between the elements along the X-axis and Z-axis.

The transmission vector from the first reflecting element to the UAV $m$ is represented as $\Delta \bar{r}_{m}(n)=\bar{l}_{m}(n)-\tilde{l}_1=(\bar x_m(n)-a,\bar y_m(n),\bar z(n)-c)^T$. The difference vector between the first reflecting element and the $k$-th reflecting element is defined as $\Delta \tilde{r}_{k}=\tilde{l}_k-\tilde{l}_1=((k_x-1)\delta_x,0,(k_z-1)\delta_z)^T$. Accordingly, for signals transmitted to the $m$-th UAV through the IRS, the phase difference between the signal reflected by the first reflecting element and the signal reflected by the $k$-th element can be formulated as follows:
\begin{align}
    &\theta_{k}^m(n) =\frac{2\pi f}{c}\frac{\Delta\tilde r_k^T}{\left|\Delta \tilde{r}_k\right|}\Delta \bar r_m(n) \nonumber \\
     &= \frac{2\pi f}{\left|\Delta \tilde r_k\right|c}\bigg((\bar x_m(n) - a)(k_x-1)\delta_x + (\bar z(n) - c)(k_z-1)\delta_z \bigg ). \nonumber
\end{align}

Similarly, the transmission vector from the first reflecting element of the IRS to user $u$ can be defined as $\Delta \hat{r}_u(n)=\hat{l}_u(n)-\tilde{l}_1=(\hat x_u(n)-a, \, \hat y_u(n),\, -c)^T$ and the phase difference between the signal sent to the user by the first reflecting element and the signal sent by the $k$-th element can be formulated as follows:
\begin{align*}
    &\nu_{k}^u(n) =\frac{2\pi f}{c}\frac{\Delta \tilde r_k^T}{\left| \Delta \tilde r_k\right|}\Delta \hat r_u(n) \nonumber\\
    &= \frac{2\pi f}{\left|\Delta \tilde r_k\right|c}\bigg((\hat x_u(n)-a)(k_x-1)\delta_x-c(k_z-1)\delta_z \bigg). \nonumber
\end{align*}

The cascaded channel gain of the UAV-IRS-UED connection can be defined as:
\begin{align}
g_{u,m}(n)&=\left(\frac{c}{8\sqrt{\pi^3} f d^{\prime}_{u,m}(n)}\right) \cdot \nonumber\\
&\exp\left(\frac{-j2\pi f d^{\prime}_{u,m}(n)}{c}+\frac{-K(f)d^{\prime}_{u,m}(n)}{2} \right), \nonumber
\end{align}
where $d^{\prime}_{u,m}(n)=\hat{d}_{u}(n)+\bar{d}_{m}(n)$. $\hat{d}_{u}(n)=||\Delta \hat r_u(n)||_2$ denotes the distance between user $u$ and the first reflector of IRS and $\bar{d}_m(n)=||\Delta \bar r_m(n)||_2$ represents the distance between the UAV $m$ and the first reflector of IRS. Finally, the channel gain of the UAV-IRS-UED data transmission is defined as:
\begin{equation}
\hat{g}_{u,m}(n)=g_{u,m}(n) \, \bar{\vec{e}}_{m}(n)^T \, \bm{\Phi}(n) \, \hat{\vec{e}}_{u}(n),
\end{equation}
where $\bm{\bar{e}}_{m}(n)=(\exp(j\theta_{1}^{m}(n)),\cdots,\exp(j\theta_K^{m}(n)))^T$, $\hat{\vec{e}}_{u}(n)=(\exp(j\nu_1^{u}(n)),\cdots,\exp(j\nu_K^{u}(n)))^T$, and $\vec{\Phi}(n)=diag(\exp(j\phi_1(n)),\cdots,\exp(j\phi_K(n)))$ is diagonal matrix of IRS phase shifts, where $\phi_k(n)$ is the phase shift of the $k$-th reflecting element. 

\subsection{System Energy Consumption}
\label{sec:sys_model_eng}
In this section, we formulate the energy consumed in the MEC system. The energy cost within the system consists of two parts: (i) the energy consumed by processing user tasks on UEDs and (ii) the energy consumed by processing user tasks on UAVs. At a given time frame $n$, let us consider user $u$ with its corresponding task denoted as $\Psi_{u}(n) = \{D_{u}(n), T_{u}(n), C_{u}(n)\}$. Here, $D_{u}(n)$ represents the size of the data, $T_{u}(n)$ represents the tolerable latency, and $C_{u}(n)$ represents the CPU cycles required to process the task. If the task is processed on the user's UED (i.e. $\beta_{u,M+1}(n)=1$), the energy consumed can be defined as:
\begin{equation}
    E^{local}_{u}(n) = t_u^{local}(n)\cdot p_u, \nonumber
\end{equation}
where $p_u$ represents the energy consumed by the UED per CPU clock and $t_u^{local}(n)$ denotes the time required for processing the user's task (measured in CPU clock):
\begin{equation}
    t^{local}_{u}(n) =\frac{C_u(n)}{Z_u}, \nonumber
    \label{local}
\end{equation}
where $Z_u$ refers to the CPU clock speed of the UED. It is assumed that both $Z_u$ and $p_u$ remain constant over time.

If user $u$'s task is processed on UAVs (i.e., $\sum_{m \in \mathcal{M}}\beta_{u,m}(n)=1$), the energy consumed during this process can be divided into two parts: (i) the energy consumed for uploading the task to UAVs and (ii) the energy consumed during the task processing on UAVs. The energy consumed in transmitting data from user $u$ to UAVs is defined as follows:
\begin{equation}
	E_{u}^{tran}(n) = t_{u}^{tran}(n)\cdot p^{tran}_u, \nonumber
\end{equation}
where $p^{tran}_u$ represents the energy consumed per second and $t_{u}^{tran}(n)$ denotes the transmission time (measured in second):
\begin{equation}
	t_{u}^{tran}(n) = \frac{D_{u}(n)}{\sum_{m\in\mathcal{M}}R_{u,m}(n)\cdot\mathbf{I}[\beta_{u,m}(n)=1]}\ , \nonumber
\end{equation}
where $\mathbf{I}[\beta_{u,m}(n)=1]$ is an indicator function that takes a value of $1$ if $\beta_{u,m}(n)=1$, and a value of $0$ otherwise.

Regarding the energy consumed in processing user $u$'s task on UAVs, it can be defined as:
\begin{equation}
    E^{comp}_{u}(n)=\sum_{m\in\mathcal{M}}t_{u,m}^{comp}(n)\cdot p_m\cdot\mathbf{I}[\beta_{u,m}(n)=1], \nonumber
\end{equation}
where $p_m$ represents the energy consumed by UAV $m$ per CPU clock, and $t_{um}^{comp}(n)$ denotes the number of CPU clocks required to process user $u$'s task on UAV $m$.
\begin{equation}
    t^{comp}_{um}(n)=\frac{C_u(n)}{Z_m/w_m(n)}. \nonumber
\end{equation}
In this context, $Z_m$ represents the CPU clock speed of UAV $m$, while $w_m(n) = \max(1, \sum_{u\in\mathcal{U}}{\beta_{u,m}}(n))$ denotes the workload status of UAV $m$. The workload refers to the current number of tasks being processed on UAV $m$.

Hence, the energy consumption attributed to user $u$ can be formulated as follows:
\begin{equation}
    E_{u}^{total}(n) = G \cdot \big(E_u^{tran}(n)+E_u^{comp}(n)\big) + (1 - G) \cdot E_u^{local}(n), \nonumber
\end{equation}
where $G = 1 - \beta_{u,M+1}(n)$.

The overall system energy is defined as the aggregate of the energy consumed by all users within the system:
\begin{equation}
    E^{total}(n) = \sum\nolimits_{u \in \mathcal{U}} E_{u}^{total}(n).
\end{equation}

\section{Optimization Problem}
\label{sec:optimization_model}
In the given system time frame $n \in \mathcal{N}$, our objective is to minimize the total energy consumption $E^{total}(n)$ within the MEC system, while considering various constraints. To simplify the notation, we denote the coordinates of all users and UAVs in the system as $\bm{L}(n)$, the CPU clock speed of UAVs and UEDs as $\bm{Z}(n)$, and the task information of all users as $\bm{\Psi}(n)$. We rewrite the total energy consumed in the system $E^{total}(n)$ as:
\begin{equation}
\label{eq:total_eng_cost}
E^{total}(n)\{\vec{\beta}, \vec{\phi}|\bm{L},\bm{\Psi},\bm{Z} \} = \sum_{u \in \mathcal{U}} E_u^{total}(n)\{\vec{\beta}, \vec{\phi}|\bm{L},\bm{\Psi},\bm{Z} \}    
\end{equation}
to highlight the dependent variables, where the `$(n)$' terms in $\bm{L}(n),\bm{\Psi}(n),\bm{Z}(n),\bm{\beta}(n),\bm{\phi}(n)$ are omitted for convenience. Accordingly, the optimization problem can be formulated as: 
\begin{align}
&\mathcal{P}1: \min_{\bm{\beta}(n),\bm{\phi}(n)} E^{total}(n)
\{\vec{\beta}, \vec{\phi}|\bm{L},\bm{\Psi},\bm{Z} \}\label{min}\\
&\mathbf{s.t. }\ \beta_{u,m}(n) \in \{0,1\} , \forall u \in \mathcal{U}, m \leq M+1,\tag{\theequation a}\label{sum not larger than one}\\
& \sum_{m=1}^{M+1}{\beta}_{u,m}(n)=1,\tag{\theequation b}\label{sumlimit}\\
&  0 \leq\phi_k(n)\leq 2\pi, 1\leq k \leq K,\tag{\theequation d}\label{phase}\\
&  t_u^{comp}(n)+t_u^{tran}(n)+t_u^{local}(n)\leq T_u(n),\forall u \in \mathcal{U}.\tag{\theequation f}\label{time}
\end{align}
It means that given $\{\bm{L,\Psi,Z}\}$, we want to find the offloading decision $\vec{\beta}(n)$ and the IRS phase $\vec{\phi}(n)=\{\phi_1(n),\phi_2(n),\cdots,\phi_K(n)\}$ such that the total energy consumed is minimized. The best offloading decision and the best IRS phase shifts are denoted as $\vec{\beta}^\circ(n)$ and $\vec{\phi}^\circ(n)$ respectively. Constraints \eqref{sum not larger than one} and \eqref{sumlimit} ensure that at the time frame $n$, each user is assigned only one task, which can be either allocated to one of the $M$ UAVs or executed locally on the UED. The constraint \eqref{phase} guarantees the angle of the $k$-th reflector of IRS remains within the range of 0 and $2\pi$. Lastly, constraint \eqref{time} ensures that the task of user $u$ is completed within the acceptable delay threshold $T_u(n)$.

Problem $\mathcal{P}1$ presents a formidable challenge as it belongs to the category of NP-hard mixed-integer non-linear programming (MINLP) problems. To tackle this challenge, we propose a two-step approach. For the first step, we focus on generating the offloading decision $\vec{\beta}^*(n)$. In this study, we introduce a deep learning-based offloading decision generation model capable of generating high-quality offloading decisions within milliseconds. The intricate details of this model are elucidated in Section \ref{sec:DNN}. Once the offloading decision $\vec{\beta}^*(n)$ is obtained from the offloading decision model, the subsequent step involves identifying the phase shifts $\vec{\phi}^*(n)$ for the IRS that minimize the overall system energy consumption, given the decision $\vec{\beta}^*(n)$. The optimization of IRS phase shifts is explained in detail in Section \ref{sec:WOA} and can be formulated as:
\begin{align}
\label{eq:optphi}
&\mathcal{P}2: \underset{\bm{\phi}(n)}{\min} \ E^{total}(n)\{\vec{\phi}|\bm{L},\bm{\Psi},\bm{Z},\vec{\beta}^*\} \nonumber \\
&\mathbf{s.t.}\ \nonumber \eqref{phase}.
\end{align}

\section{The IOPO Framework}
\label{sec:iopo}

\begin{figure*}[t]
	\centering
	\includegraphics[width=\textwidth]{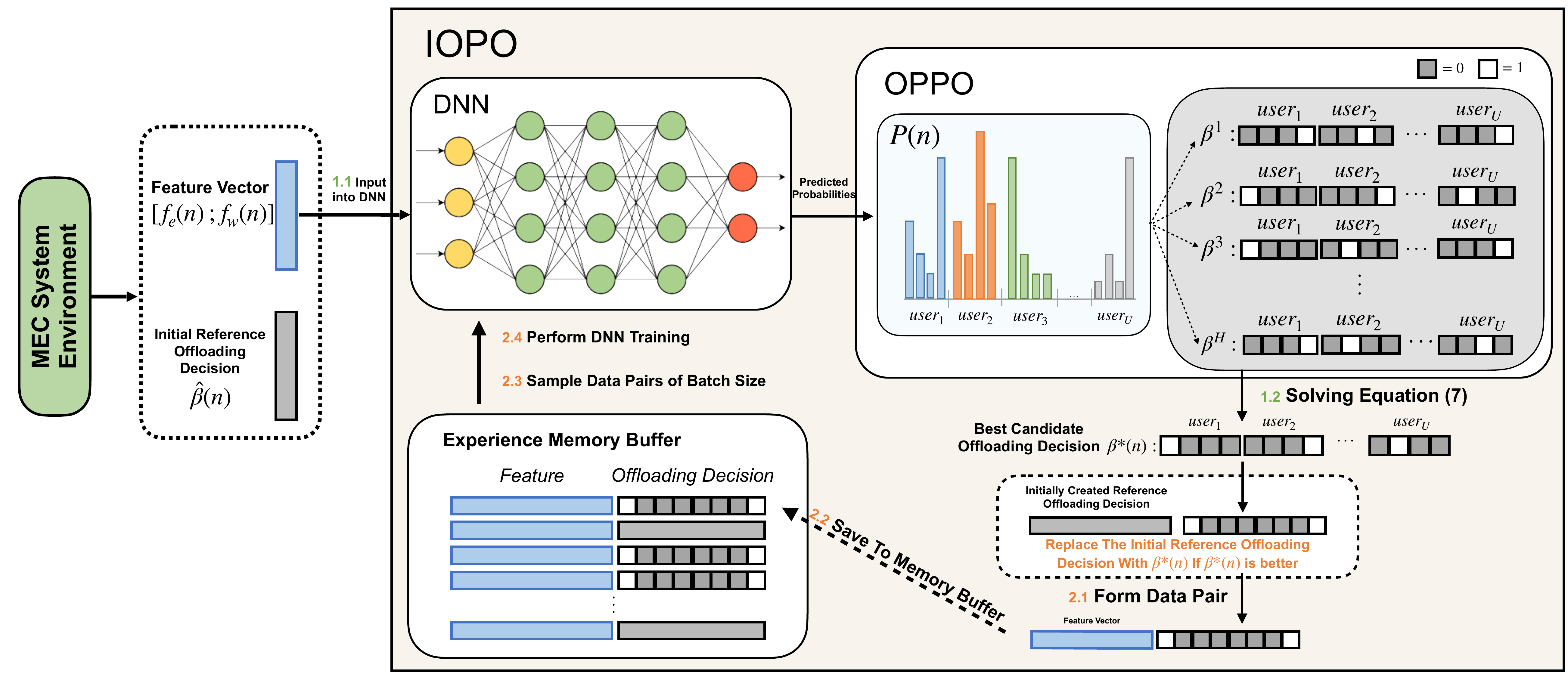}
	\caption{{The structure of the proposed IOPO Framework.}}
	\label{dnn}
\end{figure*}

\subsection{IOPO Framework Overview}
The proposed Iterative Order-Preserving Policy Optimization (IOPO) Framework, as illustrated in Figure \ref{dnn}, comprises two alternating stages: (i) \textbf{offloading decision generation} and (ii) \textbf{offloading policy update}. In the offloading decision generation stage, a deep neural network (DNN) offloading decision prediction model denoted as $f_{\theta}$ is utilized to predict an energy-efficient task offloading allocation. For the $n$-th system time frame ($n \in \mathcal{N}$), the DNN takes the input feature $[f_e(n); f_w(n)]$ constructed based on the status of system environment, and outputs a probability matrix $\bm{\mathcal{P}}(n)$, representing the probabilities of different offloading allocations that each user may adopt at time $n$. The probability matrix is then quantized into $H$ candidate offloading decisions within the Order-Preserving Policy Optimization (OPPO) unit. Among these candidate decisions, the one yielding the lowest system energy cost is selected as the predicted offloading decision for the current time frame, denoted as $\bm{\beta}^*(n)$. Subsequently, the generated offloading decision $\bm{\beta}^*(n)$, along with the corresponding input feature vector, are stored in the experience memory buffer for subsequent DNN training. 

In the offloading policy update stage, a batch of training samples is randomly selected from the memory buffer to train the DNN $f_{\theta}$, resulting in the update of DNN parameters $\theta$. The updated DNN is then utilized to produce offloading decisions in the subsequent system time frames. Detailed descriptions of these two stages are provided in the following subsections.

\subsection{Offloading Decision Generation}
\label{sec:DNN}
At a system time frame $n \in \mathcal{N}$, the input to DNN is a feature vector $\left[f_e(n); f_w(n)\right]$ formed by concatenating two distinct feature vectors: $f_e(n)$ and $f_w(n)$, where `$[\cdot;\cdot]$' denotes the vector concatenation operator. The first feature vector $f_e(n) \in \mathbb{R}^{(M+1)\times U}$ represents the energy costs associated with each of the $U$ users and their $M+1$ offloading options. The second feature vector $f_w(n) \in \mathbb{R}^{M}$ encodes the CPU clock speed of $M$ UAVs. The two feature vectors are concatenated to form the DNN input feature vector, which possesses a shape of $(M+1)\times U + M$. The DNN offloading decision model $f_{\theta}$ with parameters $\theta$, is a multilayer perceptron (MLP) consisting of an input layer, six hidden layers, and an output layer. The activation function employed in both the input and hidden layers is the hyperbolic tangent (Tanh) function, while the softmax function is utilized in the output layer. In order to enhance the model's generalization capability and mitigate the potential overfitting issue, a dropout layer \cite{dropout} is incorporated between each pair of consecutive hidden layers.

Given the input feature $\left[f_e(n); f_w(n)\right]$, the DNN predicts a probability matrix $\bm{\mathcal{P}}(n) = \{p_{u,m}(n)\ |\ p_{u,m}(n) \in \left[0, 1\right], u \in \mathcal{U}, m \in \{1, 2,\cdots, M + 1\}\}$. Each element in the matrix holds a value ranging from 0 to 1, and the matrix has a dimension of $U \times (M + 1)$. The probability matrix $\bm{\mathcal{P}}(n)$ signifies the probability of different offloading allocations that each user may adopt at the system time $n$. Specifically, the $p_{u,m}(n)$ denotes the probability that user $u$ offloads its task to UAV $m$, while $p_{u,M+1}(n)$ denotes the probability that user $u$ is assigned to execute the task locally on its UED. This process can be mathematically formulated as follows:
\begin{equation}
   \bm{\mathcal{P}}(n) = f_{\theta}\bigg(\left[f_e(n); f_w(n)\right]\bigg). \nonumber
\end{equation}
The next step is to transform the probability matrix $\bm{\mathcal{P}}(n)$ into the offloading decision matrix $\vec \beta(n)$. To accomplish this, we first feed the probability matrix into a novel Order-Preserving Policy Optimization (OPPO) unit, where $H$ candidate offloading decisions are generated based on the DNN output. Then, the candidate offloading decision with the minimum energy cost is chosen from this set of $H$ decisions to serve as the predicted offloading matrix $\vec {\beta}^*(n)$.

The OPPO unit is derived from the order-preserving optimization method proposed in \cite{2020DROO}. The original order-preserving algorithm generates a set of $H$ candidate offloading decisions, where the dissimilarity between any two candidate decisions is maximized. This approach promotes diversity among the candidate solutions, thereby increasing the chance of identifying the optimal decision. However, the order-preserving method described in \cite{2020DROO} is specifically designed for systems that consist of a single MEC infrastructure. As the proposed MEC system consists of multiple UAVs and users, the original approach is not suitable. Hence, we modify the order-preserving optimization algorithm to align with our system configuration, resulting in the modified approach referred to as OPPO. Specifically, given the DNN predicted probability matrix $\vec{\mathcal{P}}(n) \in \mathbb{R}^{U \times (M+1)}$, where $U$ represents the number of users and $M$ denotes the number of UAVs in the system, OPPO generates a set of $H$ candidate offloading decisions, where the hyper-parameter $H$ is a positive integer chosen from the range of $\{1, 2, \cdots, U \times (M+1)\}$.

The first candidate offloading decision $\vec{\beta}^1$ can be obtained through the following procedure. For the $u$-th row of $\bm{\mathcal{P}}(n)$, we identify the index of the highest probability within that row using $z_0 = \underset{z \in \{1, 2, \cdots, M+1\}}{\arg \max} p_{u,z}$. Subsequently, we set $\beta^1_{u, z_0}$ to 1, while assigning 0 to the remaining $M$ elements within that row. Mathematically, this process can be expressed as follows:
\begin{align}
\beta^1_{u,m} =  \begin{cases}
                          1 & m=z_0\ \textrm{and}\ p_{u,m} > \mathcal{T}_0,  \nonumber \\
                          0 & otherwise.
                \end{cases}
\end{align}
where $\mathcal{T}_0 = 1/(M+1)$. To generate the remaining $H-1$ offloading decisions, we begin by arranging all $U \times (M+1)$ elements of $\mathcal{P}(n)$ in ascending order based on their distances from $\mathcal{T}_0$. This sorted matrix is denoted as $\mathcal{T} = \{p'_{1,1}, p'_{1,2}, \cdots, p'_{U,M+1}\}$. Here, the element $p'_{i,j}$ becomes the $h$-th threshold denoted as $\mathcal{T}_h$, where $h = (i-1) \cdot (M+1) + j$, and $i$ and $j$ represent the row and column indices of $p'_{i,j}$, respectively. For instance, $\mathcal{T}_1 = p'_{1,1}$ corresponds to the probability element with the smallest distance to $\mathcal{T}_0$. Subsequently, the $h$-th offloading decision, denoted as $\vec{\beta}^h$ (where $h\in\{2, 3, \cdots, H\}$), is defined according to three generation rules.

The first generation rule states that for the $u$-th row of $\vec{\mathcal{P}}(n)$, if $\mathcal{R}1=\{(u, z_1)\ |\ p_{u,z_1} > \mathcal{T}_{h-1}, z_1 \in \{1, 2, \cdots, M+1\}\}$ is not an empty set, then we assign $\beta^h_{u,z_1}=1$, while setting the remaining $M$ values to 0. Mathematically, this can be expressed as:
\begin{align}
\beta_{u,m}^h =  \begin{cases}
          1 & m = z_1, \\
          0 & otherwise. \nonumber
        \end{cases}
\end{align}
If there are multiple elements in $\mathcal{R}1$, we utilize the first $(u,z_1)$ pair only and omit the remaining elements to meet the constraint (\ref{sumlimit}). In the case where $\mathcal{R}1$ is an empty set, we proceed to apply the second generation rule. Specifically, for the $u$-th row of $\vec{\mathcal{P}}(n)$, if $\mathcal{R}2=\{(u, z_2)\ |\ p_{u,z_2} = \mathcal{T}_{h-1}, p_{u,z_2} \le \mathcal{T}_0, z_2 \in \{1, 2, \cdots, M+1\}\}$ is not an empty set, we assign a value of 1 to $\beta^h_{u,z_2}$ while setting the remaining elements to 0. This can be expressed mathematically as:
\begin{align}
\beta_{u,m}^h =  \begin{cases}
          1 & m = z_2, \\
          0 & otherwise. \nonumber
        \end{cases}
\end{align}
Again, if there are multiple elements in $\mathcal{R}2$, we only utilize the first $(u,z_2)$ pair and omit the remaining elements. Lastly, in the scenario where both $\mathcal{R}1$ and $\mathcal{R}2$ are all empty, we employ the third generation rule, whereby the task is assigned to be executed locally:
\begin{align}
\beta_{u,m}^h =  \begin{cases}
          1 & m = M+1, \\
          0 & otherwise. \nonumber
        \end{cases}
\end{align}
Upon completion of the OPPO, we obtain a collection of $H$ candidate offloading decisions, denoted as $\{\beta^1, \beta^2, \cdots, \beta^H\}$. Subsequently, we identify the optimal candidate offloading decision among them, which corresponds to the one that minimizes the overall system energy cost. This process can be mathematically formulated as follows:
\begin{align}
\bm{\beta}^*(n) = \underset{\vec{\beta}^i\in \{\vec{\beta}^1,\vec{\beta}^2,\cdots,\vec{\beta}^H\}}{\arg\min} E^{total}(n)\{\bm{\beta}^i,f_{WOA}(\bm{\beta}^i)|\bm{L},\bm{\Psi},\bm{Z} \},
\label{beta_star}
\end{align}
where $E^{total}$ is Equation \ref{eq:total_eng_cost} and $f_{WOA}(\cdot)$ corresponds to the WOA method for producing optimized IRS phase shifts (introduced in Subsection \ref{sec:WOA}). 
Please be noted that, as the OPPO unit can generate $H$ candidate offloading decisions based on the DNN output, it can also be perceived as an effective solution searching unit, in which offloading decisions with low energy costs are discovered. Throughout the execution of IOPO, OPPO continuously explores offloading decisions that are more energy-efficient. These newly discovered offloading decisions are subsequently utilized in the offloading policy update procedure to update the DNN parameters $\theta$. 

After obtaining the predicted offloading decision $\bm{\beta}^*(n)$, we employ the function $\bm{\phi}^*(n)=f_{WOA}(\bm{\beta}^*(n))$ to compute the optimized IRS phase shifts $\bm{\phi}^*(n)$. By substituting $\bm{\beta}^*(n)$ and $\bm{\phi}^*(n)$ into Equation \ref{eq:total_eng_cost}, we can evaluate the energy cost of the system. However, in order to address $\mathcal{P}1$, it is imperative for the predicted offloading decision $\bm{\beta}^*(n)$ to align with, or at least closely approximate, the optimal offloading decision $\bm{\beta^{\circ}}(n)$ (i.e. $\bm{\beta^*}(n) = \bm{\beta^\circ}(n)$ or $\bm{\beta^*}(n) \approx \bm{\beta^\circ}(n)$). To achieve this alignment, it is necessary to implement an offloading policy update procedure, which enables the DNN to learn to accurately generate desired offloading decisions. Furthermore, the desired offloading decisions utilized in DNN training should also be gradually improved as the execution of IOPO. As a result, the offloading decisions predicted by the IOPO framework, which are derived from DNN outputs, exhibit a gradual improvement and ultimately converge towards optimal offloading decisions.

However, during the initial stages of the IOPO execution, the DNN is not yet adequately trained. As a result, the predicted offloading decision $\bm{\beta}^*(n)$ may exhibit poor quality. Learning from these low-quality offloading decisions could hinder the convergence towards optimal offloading decisions, particularly in systems with a substantial number of UAVs and users (wherein a poorly performing DNN finds it challenging to predict the optimal decision among a total of $(M+1)^{U}$ possible offloading decisions, with $M$, $U$ denoting the number of UAVs and the number of users within the system). To address this issue and expedite the convergence process, an intuitive approach is to provide a favorable starting point for the DNN to learn. Hence, we introduce an initial reference offloading decision $\hat{\bm{\beta}}(n)$ with high quality (the generation of this initial reference offloading decision is elaborated in Section \ref{sec:exp_iopo_execution}). At the early stages of the IOPO execution, $\hat {\bm{\beta}}(n)$ may exhibit lower energy cost compared to $\bm{\beta}^*(n)$, thereby enabling faster convergence toward the optimal offloading decisions when learning from $\hat {\bm{\beta}}^*(n)$. As the IOPO execution progresses, the DNN gradually improves, and the predicted offloading decision $\bm{\beta}^*(n)$ based on the DNN output can surpass the initial reference offloading decision. Consequently, we compare the predicted offloading decision $\bm{\beta}^*(n)$ with the initially provided reference offloading decision $\hat{\vec{\beta}}(n)$. If the MEC system achieves lower energy costs with $\bm{\beta}^*(n)$ compared to $\hat{\bm{\beta}}(n)$, we update the reference offloading decision to $\bm{\beta}^*(n)$ (i.e., $\hat {\bm{\beta}}(n) = \bm{\beta}^*(n) \nonumber$). This ensures that the DNN can always learn from high-quality offloading decisions.

Subsequently, we maintain a memory buffer with limited capacity. At the $n$-th time frame, a new training data sample $([f_e(n);f_w(n)], \hat {\bm{\beta}}(n))$ is added to the memory buffer. When the memory buffer is full, the newly generated data sample replaces the oldest one.

\subsection{Offloading Policy Update}
\label{sec:model_train}
To train the DNN offloading decision model $f_{\theta}$, first, we sample a batch of data pairs, denoted by $\mathcal{B}$, from the memory buffer, where $j\in\mathcal{B}$ implies the data pair generated in $j$-th time frame, $([f_e(j);f_w(j)], \hat {\bm{\beta}}(j))$, is in this batch. Subsequently, the parameters $\theta$ of the DNN are updated to minimize the average Maximum Likelihood Estimation (MLE) loss. The MLE loss for pair $j$ in the training batch $\mathcal{B}$ is defined as follows:
\begin{equation}
    \ell(j)= -\sum_{u=1}^U \sum_{m=1}^{M+1} \hat {\beta}_{u,m}(j)\log 
  \left( p(\hat {\beta}_{u,m}(j)|\left[f_e(j);f_w(j)\right],\theta) \right), \nonumber
\end{equation}
where $\hat {\beta}_{u,m}(j)$ refers to the reference allocation decision of the data pair $j\in\mathcal{B}$ and $[f_e(j); f_w(j)]$ is the input feature associates with the data pair $j\in\mathcal{B}$. The average MLE loss for the given training batch is formulated as:
\begin{equation}
    \mathcal{L}(\mathcal{B}) = \frac{1}{|\mathcal{B}|} \sum_{j\in\mathcal{B}} \ell(j), \nonumber
\end{equation}
where $|\mathcal{B}|$ denotes the batch size. The parameter $\theta$ is updated using the Adam optimizer \cite{adam} and is updated every $\lambda$ IOPO execution step. By minimizing $\mathcal{L}$, the IOPO-predicted offloading decisions are refined progressively and eventually align with optimal offloading decisions (demonstrate in experiment \ref{sec:exp_optimal_align}). With the optimal offloading allocations produced and the optimal phase shifts obtained using the WOA algorithm (introduced in Subsection \ref{sec:WOA}), problem $\mathcal{P}1$ can be solved. The pseudo-code of IOPO is presented in Algorithm \ref{alg:iopo}.
\begin{algorithm}[!t]
\label{alg:iopo}
\LinesNumbered 
\SetAlgoLined
\SetKwInOut{Input}{Input}\SetKwInOut{Output}{Output}
\Input{Input feature $f(n) = [f_e(n); f_w(n)]$ at each time frame $n$, and an initial reference offloading decision $\hat {\vec{\beta}}(n)$.}
\Output{Final Offloading decision $\hat {\vec{\beta}}(n)$ and the best IRS phase shifts for each time frame $n$.}
Randomly initialize parameters $\theta$ of DNN $f_{\theta}$ and empty the memory buffer.\;
\For{$n=1,2,\dots,N$}{
Compute the DNN probability matrix: $\bm{\mathcal{P}}(n) = f_{\theta}(\left[f_e(n);f_w(n)\right])$\;

Feed $\bm{\mathcal{P}}(n)$ into OPPO, where $\bm{\mathcal{P}}(n)$ is quantized into $H$ candidate offloading decisions\;

Select the best candidate decision $\bm{\beta}^*(n)$ using Equation \ref{beta_star}\;

Obtain the best IRS phase shifts $\bm{\phi}^*(n)$ using $\bm{\phi}^*(n) = f_{WOA}(\bm{\beta}^*(n))$ as shown in Sec. \ref{sec:WOA}\;

\If{$\bm{\beta}^*(n)$ is better than the initially provided reference offloading decision $\hat {\vec{\beta}}(n)$}{
    $\hat{\bm{\beta}}(n) = \bm{\beta}^*(n)$ \;
}

Update the memory buffer by adding $\big(f(n), \hat {\bm{\beta}}(n)\big)$\;

\If{$t \bmod \lambda = 0$}{
Randomly sample a batch $\mathcal{B}$ from the memory buffer as $\{(\left[f_e(j);f_w(j)\right],\hat{\bm{\beta}}(j)) \mid {j} \in \mathcal{B}\}$\;

Train the DNN on $\mathcal{B}$ and update $\theta$ using the Adam optimizer\;
}
}
\caption{The execution of the IOPO framework.}
\end{algorithm}

\subsection{IRS Phase Shifts Optimization}
\label{sec:WOA}
Given the offloading decision $\vec{\beta}^*(n)$, the determination of the optimal IRS phase shifts shown as Problem $\mathcal{P}2$ is a non-convex optimization problem. To address this, we follow \cite{2022Park} to employ the Whale Optimization Algorithm (WOA) \cite{2016whale}. WOA is commonly employed to tackle optimization problems such as resource allocations in wireless networks and beyond \cite{2020WOA}. In our approach, the WOA algorithm $\bm{\phi}^*(n)=f_{WOA}(\vec{\beta}^*(n))$ takes an offloading decision $\bm{\beta}^*(n)$ as input and produces the best IRS phase shifts $\bm{\phi}^*(n)$ through $\mathcal{E}= \{1, 2, \cdots, E\}$ evolution rounds, where the hyper-parameter $E$ determines the total number of evolution rounds. Initially, the whale population is represented as $\bm{\phi}'(0)=\{\bm{\phi}'_1(0), \bm{\phi}'_2(0),\cdots, \bm{\phi}'_W(0)\}$, where the hyper-parameter $W$ determines the number of whales in the environment. The $j$-th whale, denoted as $\bm{\phi}'_j(0)$, is a randomly generated IRS phase shift. During the $t$-th evolution round ($t \in \mathcal{E}$), the following operations are performed. Firstly, we obtain the best IRS phase shift that minimizes the system energy cost. This process can be mathematically formulated as:
\begin{equation}
    \bm{\phi}'_*(t)  = \underset{\bm{\phi}' \in \{\bm{\phi}'(t-1) \, \cup \,  \bm{\phi}'_*(t-1)\}}{\arg\min} E^{total}(n)\{\bm{\phi}'|\bm{L}, \bm{\Psi}, \bm{Z}, \bm{\beta}^*\}, \nonumber
\end{equation}
where $E^{total}_u(n)\{\cdot\}$ is Equation \ref{eq:total_eng_cost}, $\bm{\phi}'_*(t)$ denotes the global optimal phase shifts selected in the preceding $t$ iterations. In the case of $t=1$, we initialize $\bm{\phi}'_*(0)$ as an empty set, since the global optimal phase shift has not been determined yet. Subsequently, the WOA algorithm employs a balanced probability of 50\% to perform either a ``spiral route'' update or a ``shrink-wrap'' update. In the event that a ``spiral route'' update is chosen, the $j$-th whale within the whale population (i.e. the $j$-th candidate IRS phase shifts) undergoes the following update procedure:
\begin{equation}
\begin{aligned}
     &\vec{D}= abs(\bm{\phi}'_*(t)-\bm{\phi}'_{j}(t-1)), \nonumber \\
     &\bm{\phi}_{j}'(t)= abs(\vec{D} \cdot e^{b\cdot l_j(t)} \cdot \cos(2\pi \cdot l_j(t)) + \bm{\phi}'_j(t-1)), \nonumber
\end{aligned}
\end{equation}
where $abs(\cdot)$ denotes the element-wise absolute function, $b$ is a constant with a value of 1, and $l_j(t)$ denotes the behavior of the $j$-th whale during the $t$-th evolution, which is a random real value between $\left[-1, 1\right]$.

In the case of selecting a ``shrink-wrap'' update, an additional condition check is necessary to determine whether the whale engages in exploration or exploitation. Specifically, if the condition $abs(A_j(t)) < 1$ is satisfied, an \textbf{exploitation} step is performed. Conversely, if $abs(A_j(t)) \geq 1$, an \textbf{exploration} step is conducted. Here, $A_j(t) = a_j(t) \cdot (2 \, r_j(t) - 1)$, where $a_j(t) = 2 \cdot \left(1 - \frac{t}{E}\right)$ is a scalar that decreases as $t$ increases, and $r_j(t)$ is a randomly generated real value in the range of $\left[0,1\right]$. 

In the \textbf{Exploitation} phase, the update rule for the $j$-th whale can be expressed as follows:
\begin{equation}
\begin{aligned}
    &\vec{D}= abs(C_j(t) \cdot \bm{\phi}'_{*}(t) - \bm{\phi}'_{j}(t-1)), \nonumber \\ 
    &\bm{\phi}_{j}'(t)= abs(\bm{\phi}'_*(t) - A_j(t) \cdot\vec{D}), \nonumber
\end{aligned}
\end{equation}
where $C_{j}(t) = 2 \cdot r_j(t)$. In the \textbf{Exploration} phase, the update rule for the $j$-th whale can be defined as:
\begin{equation}
\begin{aligned}
     &\vec{D}= abs(C_j(t) \cdot \bm{\phi}^{rand}_j(t) - \bm{\phi}'_{j}(t-1)), \nonumber \\
     &\bm{\phi}'_{j}(t)= abs(\bm{\phi}^{rand}_j(t) - A_j(t) \cdot\vec{D}), \nonumber
\end{aligned}
\end{equation}
where $\bm{\phi}^{rand}_j(t)$ represents a randomly generated IRS phase shifts. Upon the completion of all $E$ iterations, the resulting IRS phase shifts $\bm{\phi}'_*(E+1)$ is returned as the final output of WOA.

\section{Experimental Settings}
\label{sec:exp_settings}
\subsection{Simulation Setup}
In conducted experiments, users and UAVs are confined within a rectangular area measuring 800 meters in length and 600 meters in width. Locations of users and UAVs are randomly generated within the designated area and UAVs are positioned at a fixed height of 20 meters. The CPU clock speed of MEC servers carried by UAVs, denoted as $Z_m$, is distributed between 0.08 and 0.4 GHz. In contrast, the CPU clock speed of UEDs $Z_u$ ranges from 0.04 to 0.08 GHz. The transmission frequency range from 200 to 400 GHz and the molecular absorption coefficients for THz frequencies are defined according to a reference \cite{2018absorb}. The IRS is composed of 25 reflectors, with the first element located at (4 m, 0 m, 4 m), and $K_x=5, K_z=5$. The task size of each user ranges from 32 bytes to 100 KB. The time that users finish their tasks locally is set as the acceptable delay threshold. Any processing time that is longer than this threshold fails to meet constraint (\ref{time}) and is considered as overdue.

\subsection{The Execution Of IOPO}
\label{sec:exp_iopo_execution}
We execute IOPO for $N=200,000$ system time frames, during which the DNN offloading decision model $f_{\theta}$ is trained in a supervised manner. The initial reference offloading decision is generated using the \textsc{Greedy OC} method (introduced in Section \ref{sec:compare_model}) and the training interval $\lambda$ is set to 10, indicating that the DNN parameters $\theta$ are updated every 10 IOPO execution steps. Furthermore, we utilize a batch size of 256, a dropout rate of 0.1 to mitigate overfitting, a memory buffer size of 1.5 times the batch size, and a learning rate of 0.001 in the Adam optimizer. During the execution of IOPO, we set the number of candidate decisions generated in OPPO as $H=20$. In order to guide OPPO towards identifying decisions that satisfy the no-overdue constraint (defined in Equation \ref{time}), we introduce an overdue penalty to candidate offloading decisions involving overdue users. Each overdue user adds a penalty score of 100 to the total system energy cost. This prioritizes candidate decisions without overdue users during the selection of the best candidate offloading decision. For the WOA method, the number of whales $W$ is set as 3, while the evolution round $E$ is set as 5.

Upon the completion of IOPO execution, we conducted a series of experiments to evaluate its performance compared to several offloading decision-generation baselines. These experiments are carried out over the last 1,000 system time frames and the average metrics (e.g. system energy costs, overdue statistics) are reported. To calculate the system energy costs of different methods, we first acquire a predicted offloading decision from each of the considered offloading decision models. Subsequently, we employ the WOA method denoted as $f_{WOA}(\cdot)$ to derive optimized IRS phase shifts. The optimized IRS phase shift and the obtained offloading decision are substituted into Equation \ref{eq:total_eng_cost}, yielding the total energy cost of different offloading decision generation methods.

\subsection{Comparison Offloading Decision Generation Methods}
We compare the performance of the proposed IOPO model with baseline offloading allocation approaches as follows:

\begin{itemize}
\item \textbf{\textit{Greedy Selection (Greedy)}}: This method utilizes a greedy approach to assign users to UAVs. Specifically, the algorithm iteratively selects the user with the longest local processing time and assigns it to the UAV with the fastest processing speed. After each assignment, the computational speeds of UAVs are updated based on their workload status. This process continues until the fastest UAV processing speed is slower than the slowest local computational speed among the remaining users. The remaining unassigned users finish the tasks locally.

\item \textbf{\textit{Greedy Selection with no-overdue constraint (Greedy OC)}}: Similar to the Greedy method, users are ranked based on their local processing times. However, instead of directly assigning each user to the fastest UAV, a more involved iterative process is performed. This process considers all UAVs and selects the UAV that can complete the user's task with the lowest energy cost while ensuring that the time constraints (\ref{time}) of all users on that UAV are met. If a suitable UAV cannot be found, the user is assigned to local processing.

\item \textbf{\textit{Local Computing (LOCAL)}}: All users independently process their tasks on their UEDs. No UAV resource is utilized.

\item \textbf{\textit{Optimized Random Selection (OPT RANDOM)}}: Users are randomly assigned to either local processing or UAV processing. 10 offloading decisions are randomly generated, and the decision with the lowest energy cost is selected as the final offloading decision.

\item \textbf{\textit{Optimized Random Edge Selection (OPT RANDOM w/o LOCAL)}}: Users are randomly assigned to UAVs for task processing. In this case, no user performs tasks locally. Again, 10 offloading decisions are randomly generated, and the decision with the lowest energy cost is chosen.
\label{sec:compare_model}
\end{itemize}

\section{Experimental Results}
\label{sec:exp_results}

\begin{table*}[tbp]
\renewcommand{\arraystretch}{1.2}
\centering
\caption{Overdue Statistics of Methods given different numbers of users in the system. \textit{O} Plan\% denotes the proportion of offloading decisions that contain overdue users and Avg \#\textit{O} Users denotes the average number of overdue users in overdue offloading decisions.}
\label{tab:exp_different_users_overtime_status}
\begin{tabularx}{\textwidth}{X*{2}{>{\centering\arraybackslash}X}|*{2}{>{\centering\arraybackslash}X}|*{2}{>{\centering\arraybackslash}X}}
\hline
 & \multicolumn{2}{c|}{\textsc{10 Users}} & \multicolumn{2}{c|}{\textsc{15 Users}} & \multicolumn{2}{c}{\textsc{20 Users}} \\
\textbf{Methods} & \textit{O} Plan\% & Avg \#\textit{O} Users &  \textit{O} Plan\% & Avg \#\textit{O} Users  &  \textit{O} Plan\% & Avg \#\textit{O} Users  \\ \hline
\textbf{Baselines} & & & & & & \\ \hdashline
\textsc{Local}                         & 0	        &    0	& 0	        & 0	     & 0	    & 0     \\
\textsc{Greedy (OC)}                   & 0	        &    0	& 0	        & 0	     & 0	    & 0     \\
\textsc{Greedy}                        & 81.76\%	& 1.27	& 100\%	    & 12	 & 100\%	& 12.39  \\ 
\textsc{OPT Random}                    & 82.46\%	& 3.34	& 99.94\%   & 8.83	 & 100\%	& 14.49  \\
\textsc{OPT Random}                    &            &       &           &        &          &      \\
\textsc{(w/o Local)}                   & 97.94\%	& 4.44	&100\%	    & 11.91	 & 100\%	& 17.41 \\ \hline
\textbf{Ours} & & & & & & \\ \hdashline
\textsc{IOPO }                   & 0.86\%	    & 1.36	&0.6\%	    & 1.94	 & 6.88\%	& 1.66 \\ \hline
\end{tabularx}
\end{table*}

\subsection{Model Performance Given Different Numbers Of Users}
In this experiment, we assess the proposed IOPO model in systems with varying numbers of users. The number of UAVs in systems is fixed at 3. The energy costs of offloading decisions predicted by different offloading decision models are presented in Table \ref{tab:exp_different_users_energy_costs}. It is observed that the predicted offloading decisions include users who fail to meet their acceptable delay threshold (i.e. fail to meet the constraint \ref{time}). As the ideal offloading decisions should minimize energy costs while satisfying the no-overdue constraint \ref{time}, we introduce an overdue penalty to offloading decisions containing overdue users. Specifically, each overdue user adds a penalty score of 100 to the overall system energy cost. By incorporating this overdue-penalized energy cost metric, we are able to evaluate the offloading decisions in terms of both energy costs and the occurrence of overdue users. The results presented in Table \ref{tab:exp_different_users_energy_costs} demonstrate that, in comparison to the baselines, the proposed \textsc{IOPO} model achieves the lowest overdue-penalized energy costs across all system configurations. This highlights the effectiveness of \textsc{IOPO} in generating offloading decisions that not only minimize energy consumption but also adhere to the no-overdue constraint \ref{time}.

\begin{table}[tbp]
\renewcommand{\arraystretch}{1.2}
\centering
\caption{Energy Costs of Methods Given Different Numbers of Users In the System (With Overdue Penalty = 100)}
\label{tab:exp_different_users_energy_costs}
\begin{tabular}{lccc}
\hline
\textbf{Methods} & \textbf{10 Users} & \textbf{15 Users} & \textbf{20 Users} \\
\hline
\multicolumn{4}{l}{\textbf{Baselines}} \\ \hdashline
\textsc{Local}                  & 1048.77 & 1676.27 & 2062.25 \\
\textsc{Greedy (OC)}            & 508.64  & 1011.89 & 1384.11 \\
\textsc{Greedy}                 & 451.66  & 1791.93 & 2030.92 \\
\textsc{OPT Random}             & 647.64  & 1540.31 & 2221.74 \\
\textsc{OPT Random (w/o Local)} & 737.47  & 1728.55 & 2343.66 \\ \hline
\multicolumn{4}{l}{\textbf{Ours}} \\ \hdashline
\textsc{IOPO}             & \textbf{397.72}  & \textbf{823.32}  & \textbf{1247.98} \\ \hline
\end{tabular}
\end{table}

\begin{table}[tbp]
\renewcommand{\arraystretch}{1.2}
\centering
\caption{Energy Costs of Methods Given Different Numbers of Users In the System (Without Overdue Penalty)}
\label{tab:exp_different_users_energy_costs_wo_penalty}
\begin{tabular}{lccc}
\hline
\textbf{Methods} & \textbf{10 Users} & \textbf{15 Users} & \textbf{20 Users} \\
\hline
\multicolumn{4}{l}{\textbf{Baselines}} \\ \hdashline
\textsc{LOCAL}                  & 1048.77 & 1676.27	& 2062.25 \\
\textsc{Greedy (OC)}            & 508.64  & 1011.89	& 1384.11 \\
\textsc{Greedy}                 & 347.48  & 591.92  & 791.24 \\
\textsc{OPT Random}             & 372.08  & 657.75	& 771.82 \\
\textsc{OPT Random (w/o Local)} & \textbf{301.75}  & \textbf{537.17}  & \textbf{601.82} \\ \hline
\multicolumn{4}{l}{\textbf{Ours}} \\ \hdashline
\textsc{IOPO}             & 390.18  & 819.38	& 1211.52 \\ \hline
\end{tabular}
\end{table}

To gain deeper insights into the overdue situations in offloading decisions generated by various methods, we present the overdue statistics in Table \ref{tab:exp_different_users_overtime_status}. The term \textbf{O Plans\%} represents the percentage of model-predicted offloading decisions that include overdue users, while \textbf{Avg \#\textit{O} Users} signifies the average number of overdue users within these overdue decisions. The results reveal that, except for \textsc{LOCAL} and \textsc{GREEDY (OC)}, all baseline methods generate a considerable number of offloading decisions containing overdue users. Although \textsc{LOCAL} and \textsc{GREEDY (OC)} adhere to the no-overdue constraint, they fail to fully harness UAV resources to generate energy-efficient offloading decisions (as depicted in Table \ref{tab:exp_different_users_energy_costs_wo_penalty}, wherein the overdue penalty is excluded from the system energy cost computation). Consequently, none of the baseline methods can be considered preferable. In contrast, the proposed \textsc{IOPO} framework exhibits the ability to generate offloading allocations with lower energy costs (in comparison to \textsc{LOCAL} and \textsc{GREEDY (OC)}) while significantly reducing the number of overdue users (in comparison to \textsc{GREEDY} and random methods). These findings underscore the effectiveness of the proposed methods over the baseline approaches.

\begin{table*}[!tbp]
\centering
\caption{Overdue Statistics of Methods given different numbers of UAVs. \textit{O} Plan\% denotes the proportion of offloading decisions that contain overdue users and Avg \#\textit{O} Users denotes the average number of overdue users in overdue offloading decisions.}
\label{tab:exp_different_uavs_overtime_status}
\begin{tabularx}{\textwidth}{X*{2}{>{\centering\arraybackslash}X}|*{2}{>{\centering\arraybackslash}X}|*{2}{>{\centering\arraybackslash}X}}
\hline
 & \multicolumn{2}{c|}{\textsc{3 UAVS}} & \multicolumn{2}{c|}{\textsc{4 UAVS}} & \multicolumn{2}{c}{\textsc{5 UAVS}} \\
\textbf{Methods} & \textit{O} Plan\% & Avg \#\textit{O} Users &  \textit{O} Plan\% & Avg \#\textit{O} Users  &  \textit{O} Plan\% & Avg \#\textit{O} Users  \\ \hline
\textbf{Baselines} & & & & & & \\ \hdashline
\textsc{Local}                         & 0	        &    0	& 0	     & 0	 & 0	    & 0     \\
\textsc{Greedy (OC)}                   & 0	        &    0	& 0	     & 0	 & 0	    & 0     \\
\textsc{Greedy}                        & 100\%	    & 12.39	& 100\%	 & 16.71 & 100\%    & 6.07  \\ 
\textsc{OPT Random}                    & 100\%	    & 14.49	&100\%	 &12.49	 &99.90\%	& 9.24  \\
\textsc{OPT Random}                    &            &       &        &       &          &      \\
\textsc{(w/o Local)}                   & 100\%	    & 17.41	&100\%	 &15.56	 &100\%	    & 11.52 \\ \hline
\textbf{Ours} & & & & & & \\ \hdashline
\textsc{IOPO}                    & 6.88\%	    &1.66	&6.24\%	 &1.91	 &6.80\%	& 1.86 \\ \hline
\end{tabularx}
\end{table*}

\begin{table}[t]
\renewcommand{\arraystretch}{1.2}
\centering
\caption{Energy Costs of Methods Given Different Numbers of UAVs In the System (With Overdue Penalty = 100)}
\label{tab:exp_different_uavs}
\begin{tabular}{lccc}
\hline
\textbf{Methods} & \textbf{3UAVs} & \textbf{4UAVs} & \textbf{5UAVs} \\
\hline
\multicolumn{4}{l}{\textbf{Baselines}} \\ \hdashline
\textsc{Local}                  & 2062.25          & 2078.15	     & 1779.39 \\
\textsc{Greedy (OC)}            & 1384.11          & 1194.84	     & 1009.61 \\
\textsc{Greedy}                 & 2030.92          & 2235.64	     & 1322.54 \\
\textsc{OPT Random}             & 2221.74          & 1874.64	     & 1646.52 \\
\textsc{OPT Random (w/o Local)} & 2343.66          & 2064.96	     & 1800 \\ \hline
\multicolumn{4}{l}{\textbf{Ours}} \\ \hdashline
\textsc{IOPO}             & \textbf{1247.98} & \textbf{1059.53} & \textbf{929.15} \\ \hline
\end{tabular}
\end{table}

\subsection{Model Performance Given Different Numbers Of UAVs}
In this experiment, we evaluate IOPO in systems with varying numbers of UAVs. The number of users in the system is fixed at 20 and the overdue-penalized energy costs of different methods are reported. Table \ref{tab:exp_different_uavs} illustrates the overdue-penalized energy costs resulting from offloading allocations generated by different methods. Results show that \textsc{IOPO} consistently outperforms all baseline methods across different system configurations. This underscores IOPO's ability to yield energy-efficient offloading decisions while satisfying the overdue constraint in diverse system setups. Further insights into the overdue statistics are provided in Table \ref{tab:exp_different_uavs_overtime_status}. Once again, the results affirm that IOPO surpasses the baselines \textsc{Greedy} and \textsc{Random}, while achieving comparable performance to \textsc{LOCAL} and \textsc{GREEDY (OC)} in meeting the no-overdue constraint \ref{time}.

\subsection{How Good Is The Predicted Offloading Decision Compared To The Optimal Decision?}
\label{sec:exp_optimal_align}
\begin{figure}[tbp]
    \includegraphics[width=\linewidth]{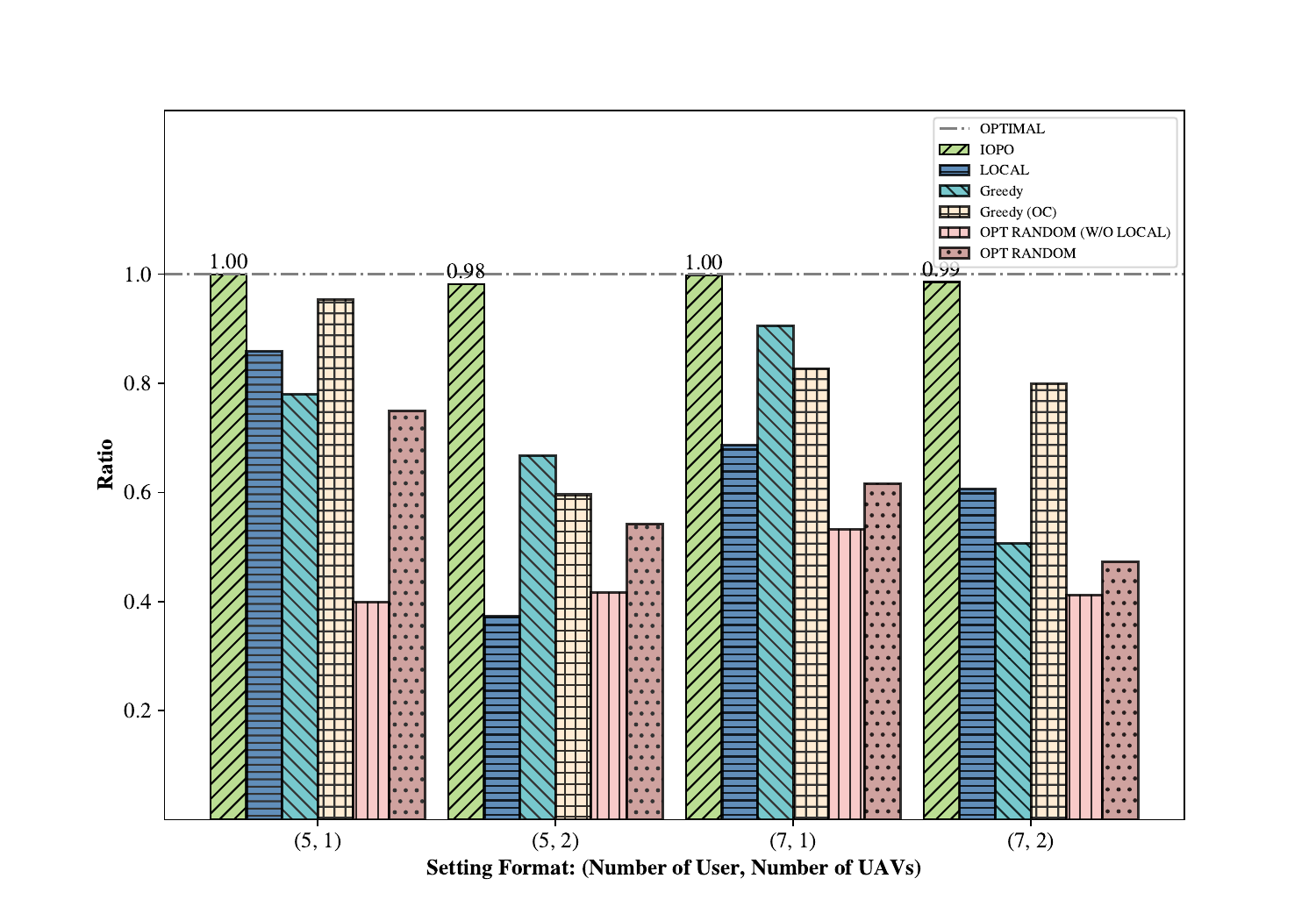}
    \caption{Average proximity ratio of methods over the last 1,000 system time frames.}
    \label{fig:optimal_during_train}
\end{figure}
In this experiment, we compare the offloading decisions predicted by IOPO with the optimal offloading decisions. Optimal offloading decisions are determined by considering all possible allocations and selecting the one that minimizes the energy cost while satisfying the no-overdue constraint. We evaluate the performance of IOPO in systems containing (5, 7) users and (1, 2) UAVs. To assess the similarity between the predicted decisions and optimal decisions, we introduce a proximity ratio. This ratio is calculated by dividing the average energy cost of optimal decisions by the average energy cost of predicted offloading decisions. An ideal scenario is indicated by a ratio of 1, signifying that the model-predicted offloading decisions perfectly match the optimal offloading decisions. A ratio smaller than 1 suggests that the energy costs of predicted offloading allocations exceed the optimal energy costs. Therefore, a ratio close to one is desirable, as it indicates a close alignment between the predicted decisions and the optimal decisions. Figure \ref{fig:optimal_during_train} demonstrates the proximity ratio of IOPO along with 5 baselines under various system settings. Notably, IOPO consistently outperforms all comparison methods, maintaining a proximity ratio close to 1 across all (user, UAV) configurations. These results substantiate that the IOPO-predicted offloading decisions can converge to optimal offloading decisions.

It should be noted that as the number of users and UAVs in the system increases, the number of possible offloading decisions grows exponentially. For instance, in a system with 5 UAVs and 20 users, the total number of potential offloading decisions amounts to $(5+1)^{20}$. This exponential growth makes it impractical to obtain optimal allocations for complex system setups within a reasonable time. Consequently, we focus the investigations on systems with a limited number of users and UAVs. While we do not present optimal solutions for intricate system setups, we observe that increasing the total number of IOPO iterations yields a further reduction in the overall system energy cost. This finding implies that for systems encompassing only a small number of users and UAVs, the IOPO model can converge towards optimal offloading decisions with a relatively small number of IOPO iterations. Conversely, for complex systems involving a larger number of users and UAVs, IOPO necessitates a greater number of iterations to approximate the optimal solution. Therefore, when confronted with systems entailing a significant number of users and UAVs, it is recommended to employ a larger number of iteration steps to attain enhanced outcomes.

\subsection{Ablation Study: How OPPO Affects IOPO Performance}

\begin{table}[tb]
\centering
\caption{Ablation Study (With Overdue Penalty = 100). \textit{O} Plan\% denotes the proportion of offloading decisions that contain overdue users and Avg \#\textit{O} Users denotes the average number of overdue users in overdue offloading decisions.}
\label{tab:ablation_study}
\begin{tabular}{lccc}
\hline
\textbf{Methods}  & \textbf{Eng Cost} & \textbf{\textit{O} Plan\%} & \textbf{Avg \#\textit{O} Users} \\
\hline
\textsc{IOPO}  & \textbf{1247.98}	& 6.88\%	& 1.66 \\
w/o OPPO  & 1408.36	& \textbf{0.94\%}	& \textbf{1.57} \\ \hline
\end{tabular}
\end{table}

This experiment investigates the influence of the proposed OPPO unit on the performance of IOPO. The experimental settings are as follows, the overdue penalty is set to 100, the number of users is set to 20, and the number of UAVs is set to 3. Table \ref{tab:ablation_study} displays the average performance of the methods over the last 1,000 system time frames. \textsc{w/o OPPO} is a variant of IOPO, in which the OPPO unit is disabled during the execution of IOPO. As OPPO is disabled in \textsc{w/o OPPO}, an alternative approach is required to quantize the DNN output probability matrix into the offloading decision matrix. To address this, at the $n$-th time frame, given the DNN predicted probability matrix $\bm{\mathcal{P}}(n) \in \mathbb{R}^{U \times (M+1)}$, for each user $u \in \mathcal{U}$, we assign a value of 1 to the offloading choice with the largest probability and a value of 0 to the remaining $M$ choices. The resulting offloading decision matrix $\bm{\beta}(n)$ satisfies constraints Equation \ref{sum not larger than one}) and Equation \ref{sumlimit}. Formally:
\begin{equation}
\label{eq:off_decision_convertion}
    \begin{aligned}
        z' = \underset{z\in\{1, 2, \cdots, M+1\}}{\arg\max} p_{u,z}, \nonumber \\
        \beta_{u,m}(n) =  \begin{cases}
          1 & m = z', \\
          0 & otherwise.
        \end{cases}
    \end{aligned}
\end{equation}

Results in Table \ref{tab:ablation_study} demonstrate that the incorporation of OPPO significantly lowers the overdue-penalized system energy cost (\textbf{Eng Cost}) when compared to the \textsc{W/o OPPO} variant. Besides, we analyze the influence of removing OPPO on the occurrence of overdue cases. The \textbf{O Plan\%} metric represents the proportion of predicted decisions that contain overdue users, while \textbf{Avg \#O Users} indicates the average number of overdue users in these decisions. Interestingly, the \textsc{W/o OPPO} variant exhibits superior performance over \textsc{IOPO} in terms of reducing the occurrence of overdue decisions and overdue users. Moreover, despite being penalized more due to overdue cases, IOPO still achieves a lower overdue-penalized energy cost over the \textsc{w/o OPPO} variant. 

The reason behind these findings can be attributed to the increasing difficulty in generating offloading allocations that effectively utilize the computational power of UAVs while satisfying the no-overdue constraint. An analysis of the decisions predicted by \textsc{w/o OPPO Training} reveals that only a small fraction of users (approximately 5 out of 20) offload their tasks to UAVs. In our experimental setup, the system comprises 3 UAVs, with each UAV capable of processing tasks for approximately 4 users while ensuring timely completion. Thus, although most decisions generated by \textsc{w/o OPPO} exhibit minimal overdue cases, they fail to fully exploit the computational capabilities of UAVs in the system. In contrast, \textsc{IOPO} gradually replaces the initial reference offloading decisions with improved decisions discovered by the OPPO unit. These improved decisions exhibit better utilization of UAV resources compared to initially provided reference decisions. Consequently, it becomes more challenging for the DNN to learn to predict these decisions accurately. The difficulty arises not only in the training of DNN but also during the generation of offloading decisions. Even a slight increase in the number of users assigned to a UAV can result in a significant number of overdue users. For instance, if a UAV can support a maximum of 4 users, but the DNN predicts assigning 5 users to that UAV, all 5 users on that UAV can become overdue users. This explains why \textsc{w/o OPPO} has fewer overdue cases and why \textsc{IOPO} achieves lower energy costs even though the predicted offloading allocations contain more overdue cases.
\begin{table}[tb]
\renewcommand{\arraystretch}{1.2}
\centering
\caption{OPPO Statistics During IOPO Execution}
\label{tab:knm_training_status}
\begin{tabular}{l|c}
\hline
\textbf{Metrics}                    & \textbf{Values} \\
\hline
\textsc{\#Improved}  & 127,966 (Iter step=200K) \\
\textsc{Eng Cost (Initial)}  & 1384.57                    \\
\textsc{Eng Cost} & 1247.98                    \\ \hline
\end{tabular}
\end{table}

Table \ref{tab:knm_training_status} presents an overview of the OPPO statistics. Throughout the IOPO execution, OPPO continually explores improved decisions that surpass the initially provided reference decisions. The metric \textsc{\#Improved} quantifies the number of IOPO predicted decisions (denoted as $\bm{\beta}^*(n)$ in Figure \ref{dnn}) that exhibit lower system energy costs compared to the initial reference decisions. Results show that a total of 127,966 improved decisions are generated during 200,000 IOPO iteration steps. During the policy update stage, the DNN can learn from these improved offloading decisions. Consequently, at the completion of IOPO execution, the overdue-penalized energy cost (\textsc{Eng Cost}) is reduced to 1247.98, in contrast to the energy cost of the initial reference offloading decisions (\textsc{Eng Cost (Initial)}), which amounts to 1384.57. It is important to note that the initial reference offloading decisions do not include any overdue users, as these decisions are derived using the Greedy method with a no-overdue constraint. Therefore, the reduction in energy cost observed in \textsc{Eng Cost} does not arise from OPPO mitigating the occurrence of overdue cases within the initial reference offloading decisions. Instead, it solely originates from OPPO's ability to discover improved offloading allocations between users and UAVs.

In summary, results demonstrate the efficacy of OPPO in generating a substantial quantity of improved offloading decisions and reducing the system energy costs.

\begin{table*}[tbp]
\centering
\caption{Model Performance and OPPO Statistics with Different DNN Complexity (Overdue penalty is set to 100 in system energy cost computation)}
\label{tab:exp_model_complexity_on_OPPO}
\begin{tabularx}{\textwidth}{X*{2}{>{\centering\arraybackslash}X}|*{2}{>{\centering\arraybackslash}X}|*{2}{>{\centering\arraybackslash}X}|*{2}{>{\centering\arraybackslash}X}|*{2}{>{\centering\arraybackslash}X}} \\ \hline
 & \multicolumn{2}{c|}{\textsc{10 Users 3 UAVS}} & \multicolumn{2}{c|}{\textsc{15 Users 3 UAVS}} & \multicolumn{2}{c|}{\textsc{20 Users 3 UAVS}} & \multicolumn{2}{c|}{\textsc{20 Users 4 UAVS}} & \multicolumn{2}{c}{\textsc{20 Users 5 UAVS}}\\

\textbf{Metrics} & Ours & Simplified & Ours & Simplified & Ours & Simplified & Ours & Simplified & Ours & Simplified \\ \hline
\textbf{Eng Cost}           & \textbf{393.34} & 424.43 & \textbf{841.49} & 912.33 & \textbf{1233.76} & 1306.16 & \textbf{1047.57} & 1118.58 & \textbf{953.45} & 1044.69 \\ 
\textbf{\#Improved}   & \textbf{146505}	& 102555	  & \textbf{143939}	& 105877	  & \textbf{126177}	& 102803	  & \textbf{122078}   & 101471	  & \textbf{115477} & 85720 \\ \hline
\end{tabularx}
\end{table*}

\subsection{Does The Initial Reference Offloading Decision Help?}

In this experiment, we study if applying initial reference offloading decisions benefits the performance of IOPO. The introduction of initial offloading decisions aims to establish a favorable starting point for training the DNN in IOPO. Without the provision of initial reference offloading decisions, the DNN may learn from suboptimal offloading decisions during the early stages of IOPO execution, thereby slowing the convergence towards optimal offloading allocations and resulting in impaired IOPO performance. This issue could become particularly pronounced when dealing with a large solution space due to the increasing difficulty in identifying high-quality offloading decisions for training the DNN. Consequently, the inclusion of initial reference offloading allocations can play a critical role in guiding the training of DNN and reducing the energy costs of IOPO-predicted offloading decisions.

\begin{figure}[tbp]
    \includegraphics[width=\linewidth]{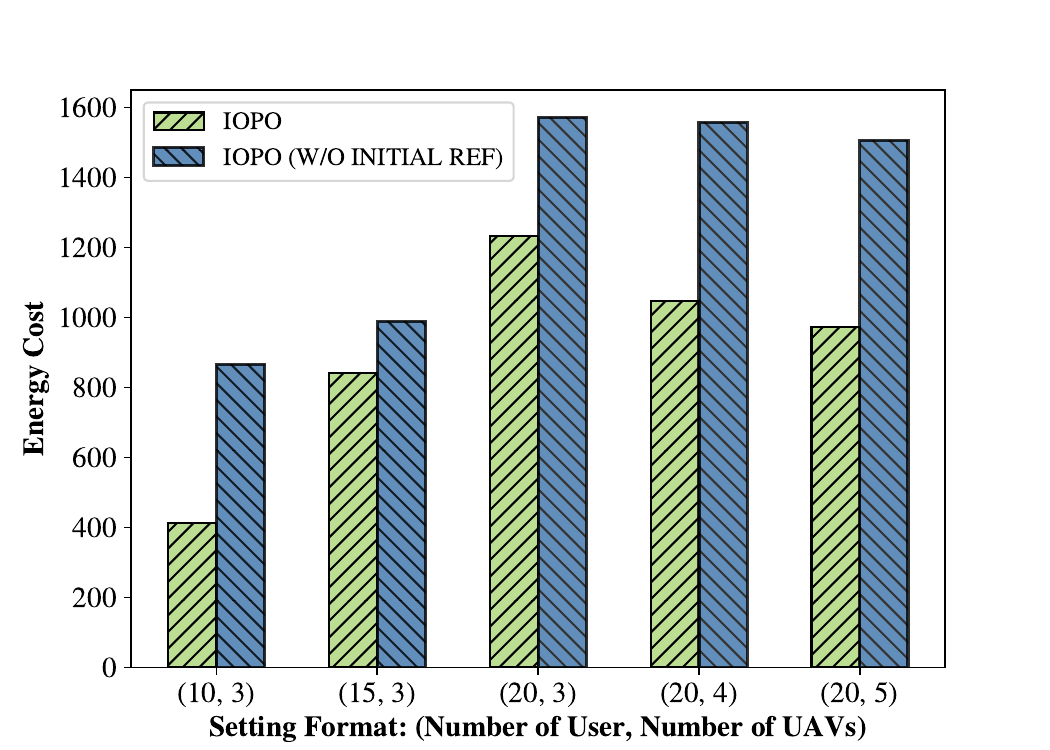}
    \caption{IOPO performance with and without utilizing initial reference offloading decisions during the training of DNN. The overdue penalty is set to 100 in system energy cost computation.}
    \label{fig:wo_initial_res}
\end{figure}

Figure \ref{fig:wo_initial_res} presents the average overdue-penalized energy costs over the last 1,000 system time frames. When the initial reference offloading decisions are not provided during DNN training, we set the predicted offloading decisions generated using Equation \ref{beta_star} as reference offloading decisions. Results demonstrate that, compared to the variant \textbf{IOPO (W/O INITIAL REF)}, in which initial reference offloading decisions are excluded in DNN training, \textbf{IOPO} can produce offloading decisions with lower energy costs. These findings align with the intuition and emphasize the significance of supplying high-quality initial reference decisions during DNN training to achieve reduced system energy consumption.

\subsection{Does DNN Complexity Affect IOPO Performance?}
In this experiment, we study the influence of DNN complexity on the performance of IOPO. Table \ref{tab:exp_model_complexity_on_OPPO} presents the performance of IOPO equipped with two DNNs: the proposed DNN (\textbf{Ours}) and a DNN with reduced complexity (\textbf{Simplified}). Compared to \textbf{Ours}, the downgraded network consists of 1 hidden layer instead of 6 and 64 hidden units instead of 256. Results indicate that the downgraded DNN (\textbf{Simplified}) exhibits higher overdue-penalized energy cost (\textbf{Eng Cost}) in all tested settings compared to the sophisticated DNN (\textbf{Ours}). This outcome can be attributed to the subpar performance of the simplified DNN in producing high-quality probability matrices. As the offloading decisions predicted by the IOPO are derived from the DNN probability matrix, sub-optimal probability matrices generated from \textbf{Simplified} result in predicted offloading decisions that incur higher energy costs. Moreover, a reduced number of improved offloading decisions discovered by OPPO (\textbf{\#Improved}) is observed in the downgraded model. These findings suggest that DNN complexity has a significant impact on the final system energy cost and the performance of OPPO searching.

\subsection{Model Analysis: Memory Buffer Size}
\label{sec:memory}

\begin{figure}[tbp]
    \includegraphics[width=\linewidth]{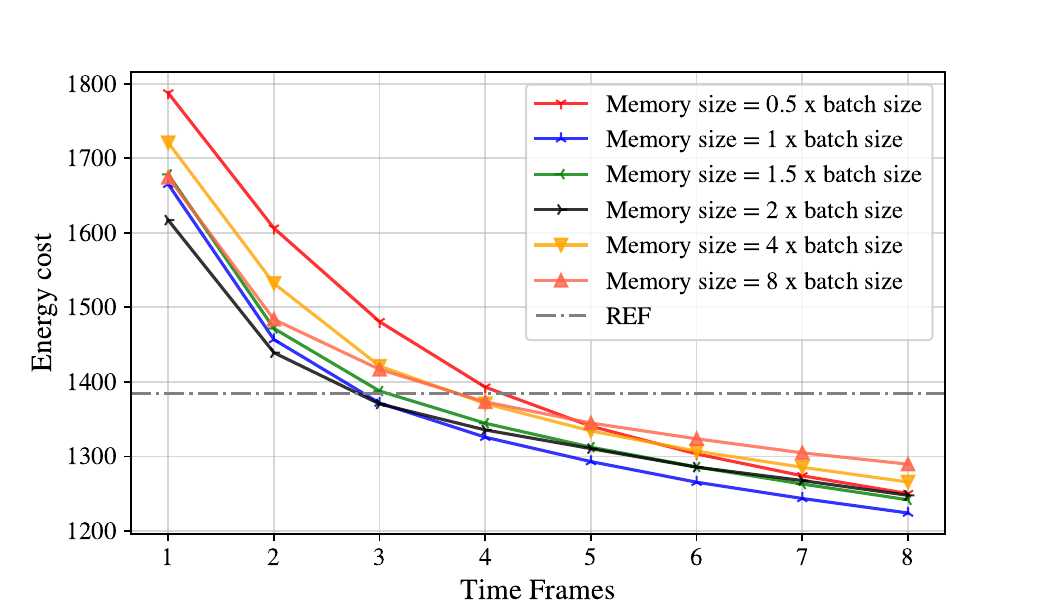}
    \caption{Impact of memory buffer size on system energy cost. Each time frame represents the average energy costs over 25,000 IOPO execution steps.}
    \label{fig:eng_cost_different_memory_size}
\end{figure}

In this experiment, we investigate the influence of memory buffer size on the performance of IOPO. The number of users in the system is set to 20, and the number of UAVs is set to 3. Figure \ref{fig:eng_cost_different_memory_size} shows the overdue-penalized energy costs of offloading decisions predicted by IOPO during the entire IOPO execution. The \textsc{REF} horizontal line represents the average energy cost of the initially provided reference offloading decisions. As depicted in Figure \ref{fig:eng_cost_different_memory_size}, IOPO with various memory sizes outperforms the REF offloading decisions as the iteration progresses. This improvement is attributed to the OPPO unit in IOPO, which can discover offloading decisions with low energy costs as the IOPO execution progresses. Moreover, IOPO with a memory size equal to the batch size demonstrates the lowest energy cost by the end of IOPO execution, compared to other memory size configurations. To provide a comprehensive understanding of the impact of memory size, Table \ref{tab:exp_memory_size} presents the average overdue-penalized energy costs (\textbf{Eng Cost}) over the last 1,000 system time frames and the number of IOPO-predicted offloading decisions that surpass the initially provided reference offloading decisions (\textbf{\#Improved}). Results indicate that the optimal IOPO performance is achieved when the memory size aligns with the batch size, with the lowest test energy cost recorded as 1232.28 and the largest number of improved allocations discovered as 131,871. These findings highlight the significance of aligning the memory size with the size of training batches for optimal IOPO performance.

When considering other memory sizes, we observe slightly higher system energy costs and smaller numbers of offloading decisions discovered compared to the optimal configuration. Additionally, as the memory size becomes larger, the overall energy cost increases. This phenomenon can be attributed to the difficulty of sampling the most recently improved offloading decisions from a substantial historical pool when training the DNN. As a result, the DNN may acquire knowledge from sub-optimal historical data, leading to compromised performance and heightened energy consumption in IOPO-predicted offloading decisions.

\begin{table}[tbp]
\centering
\caption{IOPO Performance With different Memory Sizes (With Overdue Penalty = 100)}
\label{tab:exp_memory_size}
\begin{tabular}{lcc}
\hline
\textbf{Memory Size} & \textbf{Eng Cost} & \textbf{\#Improved} \\
\hline
0.5 batch size  & 1256.82	& 121689 \\
1 batch size    & \textbf{1232.28}	& \textbf{131871} \\
1.5 batch size  & 1253.76	& 124038	 \\
2 batch size    & 1273.86	& 121413	 \\
4 batch size    & 1285.07	& 117858	 \\
8 batch size    & 1294.34	& 111396	 \\ \hline
\end{tabular}
\end{table}
\subsection{Model Analysis: Training Interval}
\begin{figure}[tb]
    \includegraphics[width=\linewidth]{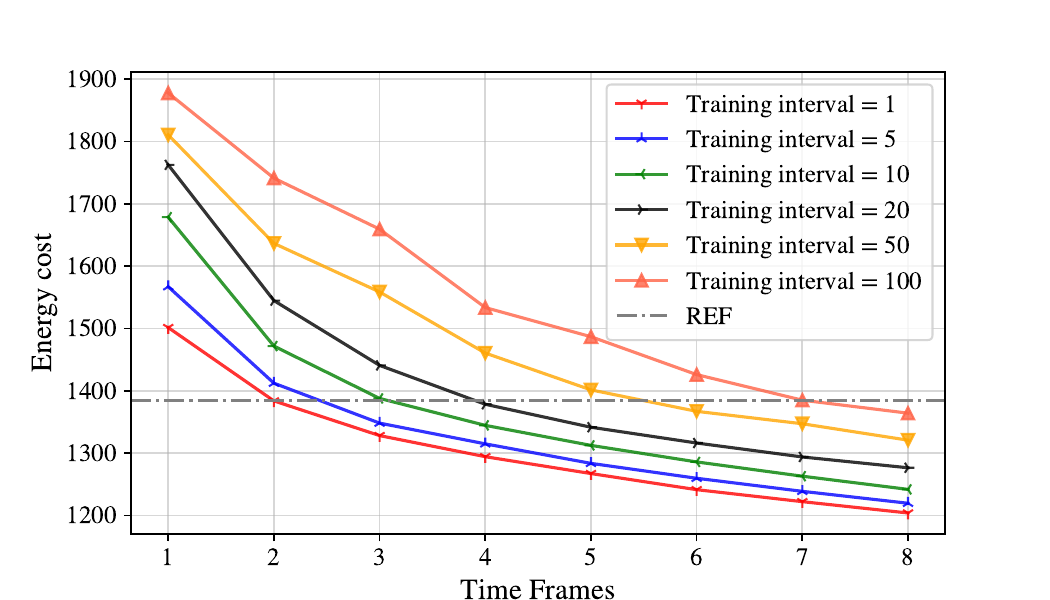}
    \caption{Impact of Training Interval size on energy cost. Each time frame represents the average energy cost over 25,000 IOPO execution steps.}
    \label{fig:eng_cost_different_TI}
\end{figure}

In this experiment, we examine the impact of the size of the training interval $\lambda$ on the performance of IOPO. The number of users in the system is set to 20 and the number of UAVs is set to 3. Figure \ref{fig:eng_cost_different_TI} illustrates the overdue-penalized energy cost of IOPO-predicted decisions and REF denotes the average energy cost of the initial reference offloading decisions.

As shown in Figure \ref{fig:eng_cost_different_TI}, IOPO with different training interval sizes (1, 5, 10) can yield offloading decisions with similar and low energy costs at the completion of IOPO execution. When the training interval size is increased to 50 and 100, the resulting decisions exhibit higher energy costs. Moreover, the energy costs of IOPO with training intervals 50 and 100 are closer to the horizontal REF line, indicating a compromised performance of the OPPO unit in discovering improved offloading decisions when the training interval is large. This is because, with large training intervals, the parameters $\theta$ of the DNN offloading decision model $f_{\theta}$ are updated less frequently. Consequently, the accuracy of the DNN is compromised, causing the predicted offloading decisions, which rely on the DNN-output probability matrix, to be impaired. 

\begin{table}[tb]
\centering
\caption{IOPO Performance With Various Training Intervals (With Overdue Penalty = 100)}
\label{tab:exp_training_interval}
\begin{tabular}{lcc}
\hline
\textbf{Training Interval} & \textbf{Eng Cost} & \textbf{\#Improved}  \\
\hline
1       & \textbf{1196.84}	& \textbf{144841}	 \\
5       & 1203.57	& 137763 \\
10      & 1253.76	& 124038	\\
20      & 1277.90    & 118099	\\
50      & 1324.63	& 86867	\\
100	    & 1370.78	& 50734	\\ \hline
\end{tabular}
\end{table}

Table \ref{tab:exp_training_interval} demonstrates that the lowest system energy cost achieved is 1196.84, and the largest number of improved decisions discovered is 144841, both obtained when the training interval is set to 1. This is because a small training interval facilitates the update of DNN parameters and the improvement of DNN performance. With the continual improvement of the DNN, there is a corresponding enhancement in the IOPO-predicted offloading allocations that depend on the DNN's performance. Subsequently, the DNN learns from these improved offloading decisions, leading to further enhancements in its own performance and a reduction in energy costs of IOPO-predicted decisions. However, it is important to note that using a smaller training interval may result in slower system speed due to the increased frequency of DNN parameter updates. If execution speed is a primary concern, it is reasonable to consider setting the training interval to 5 or 10, as these interval sizes yield energy costs that are close to the energy cost achieved with a training interval of 1.

\section{CONCLUSIONS}
\label{sec:conclusion}
In this study, we investigate the task offloading problems in a multi-user multi-UAV MEC system that integrates an IRS and operates on the 6G THz communication network. We present the modeling of the task offloading and the task processing procedure of the MEC system within the THz network and introduce IOPO, a novel deep learning-based framework designed to optimize the energy efficiency of task offloading decisions and the phase shifts of the IRS. The IOPO framework can generate satisfactory offloading decisions within milliseconds and is incorporated with a novel offloading decision-searching unit OPPO, enabling continuous search to identify improved offloading allocations. Extensive experimental results demonstrate the superiority of IOPO over baseline methods in generating energy-efficient offloading allocations and meeting task deadlines.

\section*{Acknowledgment}
This work was supported in part by the National Key R\&D Program of China under Grant No. 2022YFE0201400, the National Natural Science Foundation of China (NSFC) under Grant No. 62202055, the Start-up Fund from Beijing Normal University under Grant No. 310432104, the Start-up Fund from BNU-HKBU United International College under Grant No. UICR0700018-22, and the Project of Young Innovative Talents of Guangdong Education Department under Grant No. 2022KQNCX102.

\bibliographystyle{IEEEtran}
\bibliography{refs-bibs}

\begin{IEEEbiography}[{\includegraphics[width=1in,height=1.25in,clip,keepaspectratio]{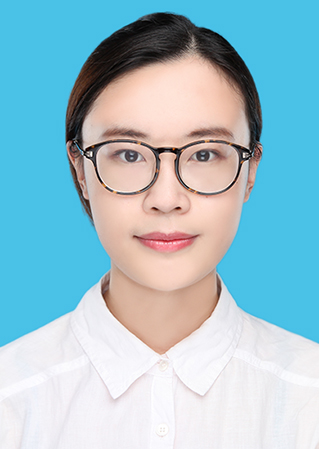}}]{Jianqiu Wu} received the M.S. degree from the Faculty of Engineering, the Chinese University of Hong Kong, in 2018. She is currently pursuing an M.Phil. degree with the Department of Computer Science, BNU-HKBU United International College, Zhuhai, China. He is supervised by Dr. Jianxiong Guo, and her research interests include reinforcement learning, mobile edge computing, and deep learning in wireless communications.
\end{IEEEbiography}

\begin{IEEEbiography}[{\includegraphics[width=1in,height=1.25in,clip,keepaspectratio]{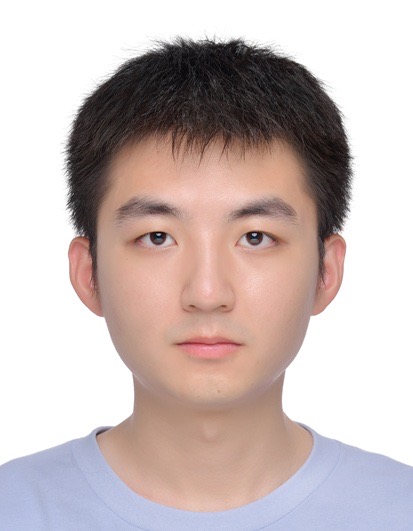}}]{Zhongyi Yu} 
received his M.S. degree from the School of Informatics at the University of Edinburgh, Edinburgh, UK, in 2022. Prior to that, he completed his B.S. degree in the Department of Computer Science at BNU-HKBU United International College, Zhuhai, China, in 2020. His research interests include reinforcement learning, natural language processing, causal inference, and efficient machine learning.
\end{IEEEbiography}

\begin{IEEEbiography}[{\includegraphics[width=1in,height=1.25in,clip,keepaspectratio]{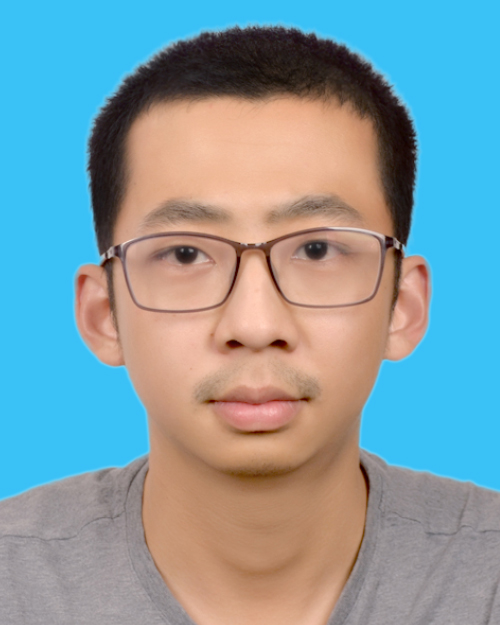}}]{Jianxiong Guo}
	received his Ph.D. degree from the Department of Computer Science, University of Texas at Dallas, Richardson, TX, USA, in 2021, and his B.E. degree from the School of Chemistry and Chemical Engineering, South China University of Technology, Guangzhou, China, in 2015. He is currently an Assistant Professor with the Advanced Institute of Natural Sciences, Beijing Normal University, and also with the Guangdong Key Lab of AI and Multi-Modal Data Processing, BNU-HKBU United International College, Zhuhai, China. He is a member of IEEE/ACM/CCF. He has published more than 40 peer-reviewed papers and been the reviewer for many famous international journals/conferences. His research interests include social networks, wireless sensor networks, combinatorial optimization, and machine learning.
\end{IEEEbiography}

\begin{IEEEbiography}[{\includegraphics[width=1in,height=1.25in,clip,keepaspectratio]{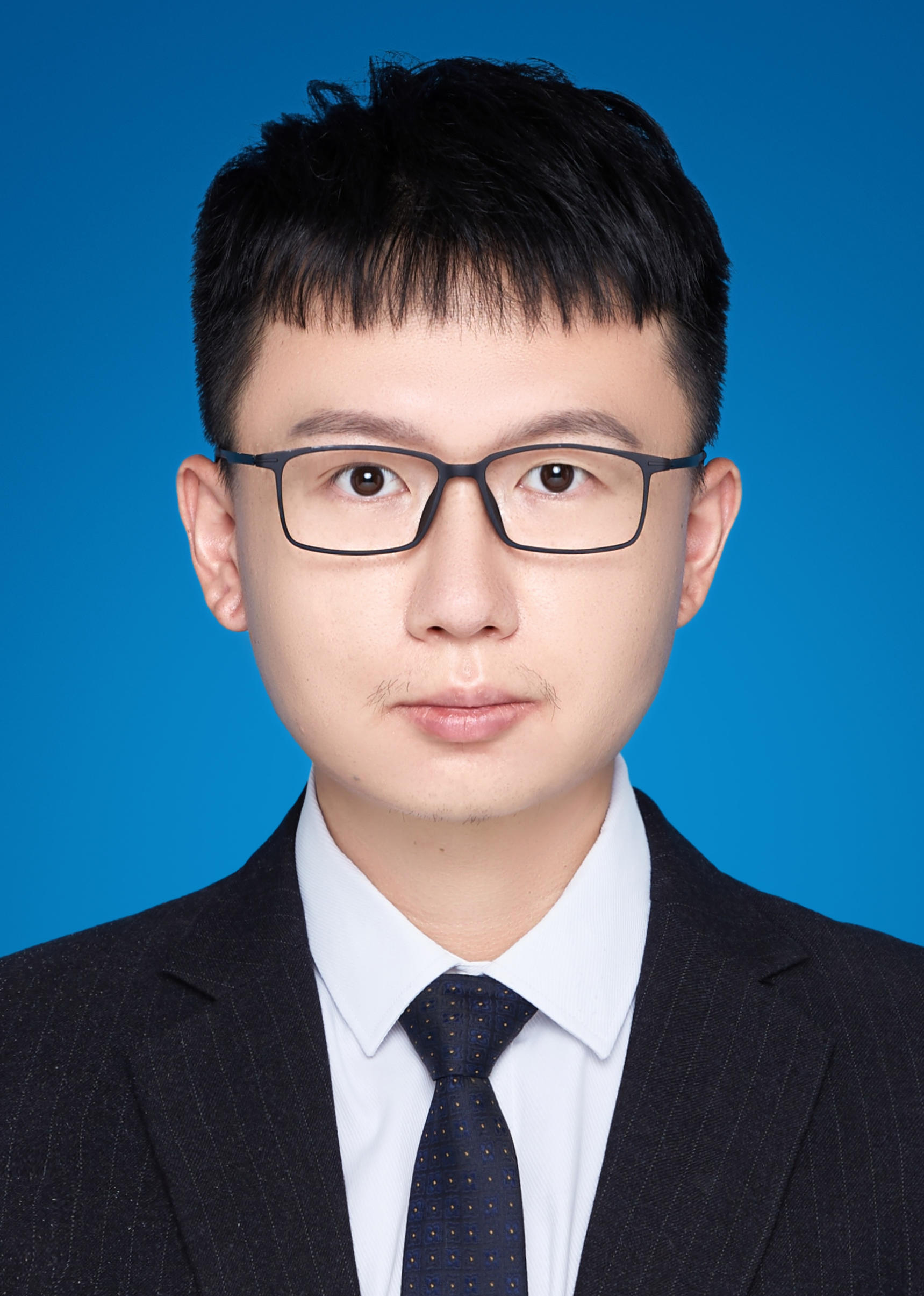}}]{Zhiqing Tang}
    received the B.S. degree from School of Communication and Information Engineering, University of Electronic Science and Technology of China, China, in 2015 and the Ph.D. degree from Department of Computer Science and Engineering, Shanghai Jiao Tong University, China, in 2022. He is currently an assistant professor with the Advanced Institute of Natural Sciences, Beijing Normal University, China. His current research interests include edge computing, resource scheduling, and reinforcement learning.
\end{IEEEbiography}

\begin{IEEEbiography}[{\includegraphics[width=1in,height=1.25in,clip,keepaspectratio]{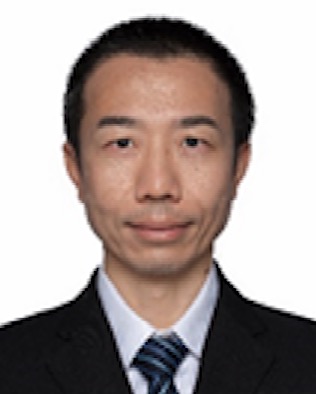}}]{Tian Wang}
	received his BSc and MSc degrees in Computer Science from the Central South University in 2004 and 2007, respectively. He received his PhD degree in City University of Hong Kong in Computer Science in 2011. Currently, he is a professor in the Institute of Artificial Intelligence and Future Networks, Beijing Normal University \& UIC. His research interests include internet of things, edge computing and mobile computing. He has 27 patents and has published more than 200 papers in high-level journals and conferences. He has more than 11000 citations, according to Google Scholar. His H-index is 53. He has managed 6 national natural science projects (including 2 sub-projects) and 4 provincial-level projects.
\end{IEEEbiography}

\begin{IEEEbiography}[{\includegraphics[width=1in,height=1.25in,clip,keepaspectratio]{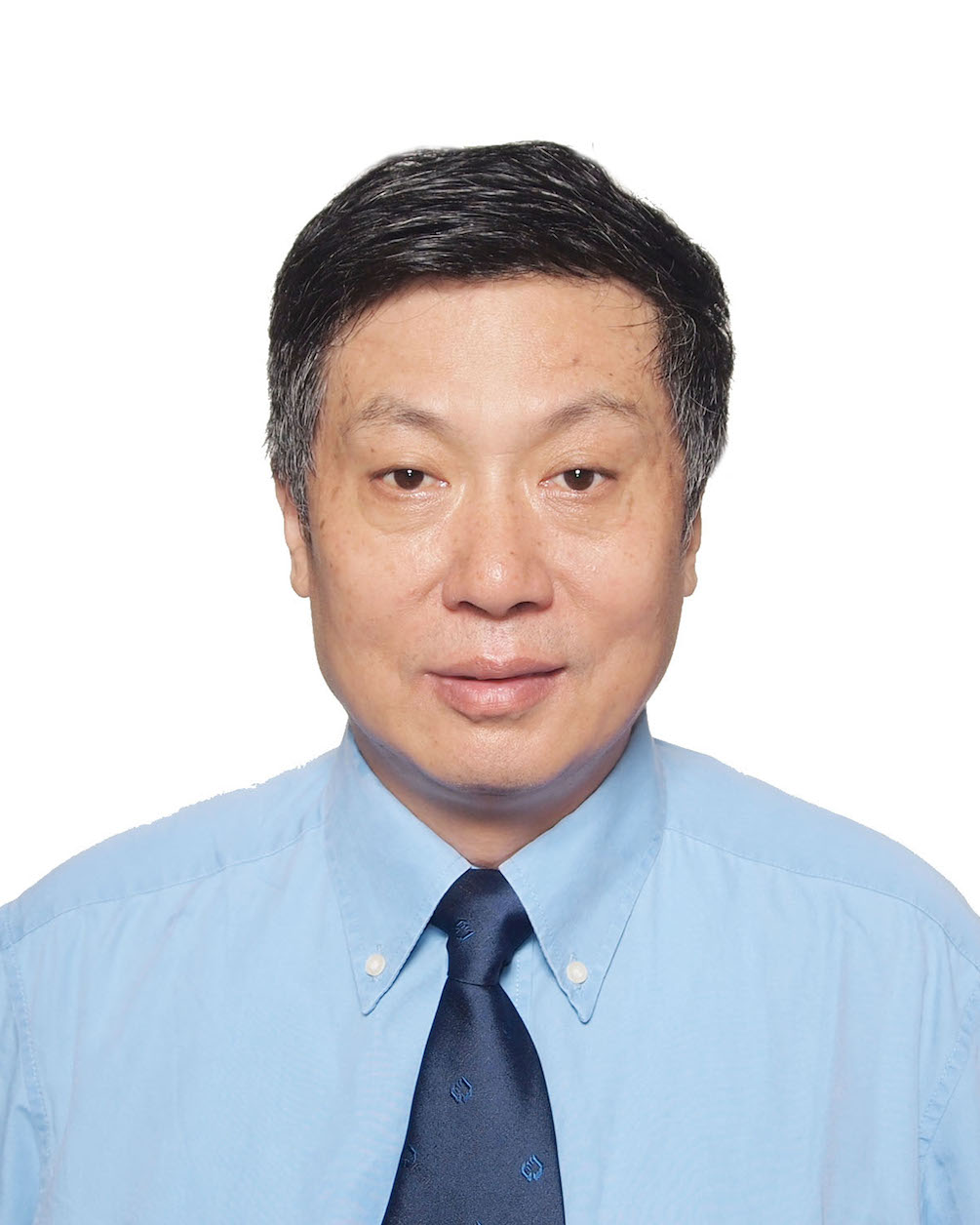}}]{Weijia Jia}
    is currently a Chair Professor, Director of BNU-UIC Institute of Artificial Intelligence and Future Networks, Beijing Normal University (Zhuhai) and VP for Research of BNU-HKBU United International College (UIC) and has been the Zhiyuan Chair Professor of Shanghai Jiao Tong University, China. He was the Chair Professor and the Deputy Director of the State Kay Laboratory of Internet of Things for Smart City at the University of Macau. He received BSc/MSc from Center South University, China in 82/84 and Master of Applied Sci./PhD from Polytechnic Faculty of Mons, Belgium in 92/93, respectively, all in computer science. From 93-95, he joined German National Research Center for Information Science (GMD) in Bonn (St. Augustine) as a research fellow. From 95-13, he worked at the City University of Hong Kong as a professor. His contributions have been recognized as optimal network routing and deployment; anycast and QoS routing, sensors networking, AI (knowledge relation extractions; NLP, etc.), and edge computing. He has over 600 publications in the prestige international journals/conferences and research books and book chapters. He has received the best product awards from the International Science \& Tech. Expo (Shenzhen) in 20112012 and the 1st Prize of Scientific Research Awards from the Ministry of Education of China in 2017 (list 2). He has served as area editor for various prestige international journals, chair, PC member, and keynote speaker for many top international conferences. He is the Fellow of IEEE and the Distinguished Member of CCF.
\end{IEEEbiography}
\end{document}